\documentclass[runningheads]{llncs}
\usepackage{graphicx}
\usepackage{cite}
\usepackage{microtype}
\usepackage{epsfig}
\usepackage[table,xcdraw]{xcolor}
\usepackage{hyperref}
\usepackage{amssymb,amsmath}
\usepackage{enumerate}
\usepackage[caption=false,position=bottom]{subfig}
\usepackage{tikz}
\usetikzlibrary{shapes.geometric, shapes.gates.logic.US,trees,arrows,shadows,positioning,calc}
\tikzset{
  frame/.style={
    rectangle, draw, 
    text width=5em, text centered,
    minimum height=2em,drop shadow,fill=yellow!60,
    rounded corners,
  },
  framext/.style={
    rectangle, draw, 
    text width=5em, text centered,
    minimum height=2em,drop shadow,fill=orange!60,
    rounded corners,
  },
  line/.style={
    draw, -latex',rounded corners=3mm,
  }
}
\tikzstyle{startstop} = [rectangle, rounded corners, minimum width=3cm, minimum height=0.95cm,text centered, draw=black, fill=orange!60]
\tikzstyle{io} = [trapezium, trapezium left angle=70, trapezium right angle=110, minimum width=3cm, minimum height=0.9cm, text centered, draw=black, fill=blue!30]
\tikzstyle{process} = [rectangle, minimum width=3cm, minimum height=0.85cm, text centered, draw=black, fill=yellow!60]
\tikzstyle{decision} = [diamond, minimum width=3.5cm, minimum height=0.5cm, text centered, draw=black, fill=green!10]
\tikzstyle{arrow} = [thick,->,>=stealth]
\tikzstyle{line2} = [thick,-]
\begin{document}
\title{Fault Trees from Data: Efficient Learning with an Evolutionary Algorithm\thanks{This research is supported by the Dutch Technology Foundation (STW) under the Robust CPS program (project 12693), the EU project SUCCESS, the Smart Industries program (project SEQUOIA 15474), and the Wallenberg AI, Autonomous Systems and Software Program (WASP) funded by the Knut and Alice Wallenberg Foundation.}}
%
%
\author{Alexis Linard\inst{1,3} \and
Doina Bucur\inst{2} \and
Mari\"{e}lle Stoelinga\inst{1,2} }
\authorrunning{A. Linard et al.}
%

\institute{Institute for Computing and Information Science \\ Radboud University, Nijmegen, The Netherlands \\
\email{a.linard@cs.ru.nl}
\and
University of Twente, Enschede, The Netherlands \\
\email{\{d.bucur,m.i.a.stoelinga\}@utwente.nl} \\
\and
KTH Royal Institute of Technology, Stockholm, Sweden
}
\maketitle              

\begin{abstract}
Cyber-physical systems come with increasingly complex architectures and failure modes, which complicates the task of obtaining accurate system reliability models.
At the same time, with the emergence of the (industrial) Internet-of-Things, systems are more and more often being monitored via advanced sensor systems.
These sensors produce large amounts of data about the components' failure behaviour, and can, therefore, be fruitfully exploited to learn reliability models automatically. 

This paper presents an effective algorithm for learning a prominent class of reliability models, namely fault trees, from observational data. 
Our algorithm is evolutionary in nature; i.e., is an iterative, population-based, randomized search method among fault-tree structures that are increasingly more consistent with the observational data.
We have evaluated our method on a large number of case studies, both on synthetic data, and industrial data. Our experiments show that our algorithm outperforms other methods and provides near-optimal results.
\keywords{fault tree induction \and safety-critical systems \and cyber-physical systems \and evolutionary algorithm.}
\end{abstract}

 \section{Introduction}
\label{sec:intro}

Reliability engineering is an important field that provides methods, tools and techniques to evaluate and mitigate the risks related to complex systems such as drones, self-driving cars, production plants, etc. Fault tree analysis is one of the most prominent technique in this field. It is widely deployed in the automotive, aerospace and nuclear industry, by companies and institutions like NASA, Ford, Honeywell, Siemens, the FAA, and many others. 

Fault trees \cite{VS02} (FTs) belong to analytical techniques for safety, security, and dependability. They are graphical models that represent how component failures arise and propagate through the system, leading to system-level failures. Component failures are modelled in the leaves of the tree as \emph{basic events}. Fault tree \emph{gates} model how combinations of basic events lead to a system failure, represented by the top event in the FT. 
The analysis of such FTs \cite{RUIJTERS201529} is multifold: they can be used to compute dependability metrics such as system reliability and availability; understand how systems can fail; identify the best ways to reduce the risk of system failure, etc. 
A key bottleneck in fault tree analysis is, however, the effort needed to construct a faithful fault tree model. FTs are usually built manually by domain experts. Given the complexity of today's systems, industrial FTs often contain thousands of gates. Hence, their construction is a very intricate task, and also error-prone, since their soundness and completeness largely depends on domain expertise. 
With the emergence of the industrial Internet-of-Things, Cyber-physical systems are more and more equipped with smart sensor systems, monitoring whether a system component is in a failed state or not. Even though such a monitoring system is often designed to detect failures during operations, their data can be very fruitfully deployed to learn reliability models. 
Such data can be crucial for the engineers to build an FT \cite{4414081}. Recent work focused on learning FTs from observational data, identifying causalities from data \cite{lift}.

\vspace{-0.1cm}
In this paper, we focus on FT generation from data, using an evolutionary algorithm (EA). 
EAs approximate stochastic learning by mimicking biological evolution, and have been successfully applied to a wide plethora of applications; examples include the scheduling of flexible manufacturing systems \cite{Geiger2006}, automata learning \cite{dupont94}, induction of Boolean functions \cite{logicalevo}, and many~more.
In our case, each stage of the EA keeps a population of candidate FTs. New fault trees are generated by mimicking biological evolution. That is, new FTs are created by reproduction (e.g., adding or deleting FT gates), crossover (e.g. swapping FT branches),  and mutation (e.g. changing an AND gate into an OR gate). In total, we have identified seven (parametric) generation rules, which are equally applied. 
Finally, we select the new population by only keeping those FTs with the best \emph{fitness}, i.e. FTs that best fit to the observational data.
We have experimentally verified the applicability of our algorithm, on synthetic data, an industrial case study, and a benchmark of FTs previously studied in the literature. Our experiments show that the algorithm is fast and accurate ($> 99\%$).
Further, we have investigated the robustness of our method to noisy data. Here we found that our EA handles noisy records.
We also developed a variation of our EA in order to take expert knowledge into account.
When domain experts partially know the structure of the FT, then the task is reduced to evolve \emph{sub}-Fault Trees, given the known \emph{skeleton} of the FT.

\vspace{-0.1cm}
Being a first step, our algorithm focuses on \emph{static} fault trees, featuring only Boolean gates. An important topic for future work is the extension to dynamic fault trees. These come with additional gates, catering for common dependability patterns like spare management and functional dependencies. Static fault trees, however, have appeal as relatively simple yet powerful formalism and are often used in practice. 
Furthermore, dynamic fault trees strongly depend on the temporal order in which failures occur, and their learning will, therefore, require more complex data, such as time series. 

This paper is organized as follows. Sections \ref{sec:sota} and \ref{sec:def} review related work on learning FTs from data as well as preliminary definitions.
We present then in Section \ref{sec:ea} our technique to infer an FT using an EA.
In Section \ref{sect-partialFT}, the variation of our EA that takes expert knowledge into account.
In Section \ref{sec:expe}, we show the results we achieved.
Finally, we discuss and conclude about further research.

\vspace{-0.2cm}
\section{Related Work}
\label{sec:sota}
\vspace{-0.2cm}

Related work on learning fault trees spans three areas of research: the synthesis of fault trees from other graphical models of the system under study; recent work on the generation of fault trees from observational data describing the system; and, since fault trees are in essence Boolean functions, literature on learning Boolean functions from observational data.

\vspace{-0.1cm}
\paragraph{Model-based synthesis.} While state-of-the-art fault tree design is often performed manually by domain specialists~\cite{kabir2017overview}, 
several methods have been proposed to synthesize FTs automatically from other models of the system~\cite{MBDA-sharvia, bozzanocompass}. 
Thus, these methods require the pre-construction of a system model in a suitable model description language, which varies with each method for FT synthesis. 
For example, the HiP-HOPS framework~\cite{HiP-HOPS-papadopoulos} synthesizes an FT from a system model describing transactions among the system components, annotated with failure information. Similar synthesis methods were developed from the AltaRica system description language, which models the causal relations between system variables and events using transitions~\cite{FT-li}. Specific system control models in the form of directed graphs have also been shown to suit the synthesis of fault trees~\cite{allen1984digraphs,henry1997computerized}, as well as Go models~\cite{zhang2015method}. 
Furthermore, system models described in the model language NuSMV also enable the synthesis of FTs. A limitation of this method is, however, the fact that the resulting FTs show the relation between top events and basic events, but do not show how failure propagates in the system via system components~\cite{bozzano2007fsap}. Static FTs can also be synthesized from models in the Architectural Analysis and Design Language AADL~\cite{li2011method}. 

As a special case, FT generation has been attempted so that the learning method includes explicit reasoning about the causal relations between events in the system. For this type of FT generation, \cite{leitner2013probabilistic} requires a probabilistic system model, from which a model-checking step obtains a set of probabilistic counterexamples. When the system is concurrent, the order of events in these counterexamples does not necessarily signify causality, so logical combinations of events are separately validated for causality. Similarly, in~\cite{fault-tree-gen-specs} a cause-effect graph (and from that, an FT) is extracted by model checking a process already modelled by a finite-state machine, against safety and liveness requirements, using failure injection.
Since model-based FT learning requires prior modelling of the system under study, these methods do not \emph{adapt} well in applications where the systems evolve and thus need to be remodelled, e.g., components are replaced, or the interactions between components change, thus changing the failure modes and their probability of occurrence.

\vspace{-0.1cm}
\paragraph{Learning causal models from data.} Supervised \emph{automated learning} of dependability models using \emph{system data}, unlike the model-based methods described above, will adapt to system change, under the assumption that all the system components remain monitored by sensors throughout their lifetime, also after a change of components. Here, we take ``learning'' to mean broadly any autonomous computational intelligence method able to infer (or even approximate) high-level models of knowledge from data. 
Causal Bayesian Networks~\cite{li2007knowledge} are standard graphical models which have been learnt from data examples. These models have straightforward translations into FTs, but are themselves NP-hard or require exponential time to synthesize accurately~\cite{BN-chickering,Kearns:1994:LBF:195613.195656}. These networks will model a limited form of causality, namely global causal relationships, rather than a sequence of causal relationships among events local to the components of a~system.

LIFT~\cite{lift} is a recent approach for learning static FTs with Boolean event variables, n-ary AND/OR gates, annotated with event failure probabilities. The input to the algorithm is untimed observational data, i.e., a dataset where each row is a single observation over the entire system, and each column records the value of a system event. All intermediate events to be included in the FT must be present in the dataset, but not all may be needed in the FT, and a small amount of noise in the dataset can be tolerated. LIFT also includes a causal validation step (the Mantel-Haenszel statistical test) to filter for the most likely causal relationships among system events, but the worst-case complexity is exponential in the number of system events in the data. Its main advantage is that of being one of the few automated FT-learning methods which validate causality.

\vspace{-0.1cm}
\paragraph{Learning Boolean formulas and classifiers from data.} Before LIFT, observational data were used to generate FTs with the IFT algorithm~\cite{madden1970generation} based on standard decision-tree statistical learning. The advantage of learning a graphical decision tree out of data is the inherent interpretability of decision-tree models and their ease of translation into other graphical models. 
Boolean formulas or networks were also machine-learnt using a similar tree-based 
method~\cite{oliveira1994learning, Kearns:1994:LBF:195613.195656}.
The classic C4.5 learning algorithm
yields a Boolean decision tree that is easily translatable into a Boolean formula by constructing the conjunction of all paths leading to a leaf modelling a True value (i.e., system failure), and then simplifying the Boolean function. The resulting models encode the same information as a decision tree (i.e., a classifier for the observational data), so lack the validation of causal relations, but are expected to preserve their predictive power about the system. This retained our attention: indeed, static FTs (in opposition to dynamic FTs, where time-dependence of events is considered) can be seen as Boolean functions. Furthermore, Boolean formulas were also machine-learnt using black-box classifiers (namely, classifiers not easily interpretable as a graphical model).
Such methods include SVMs, Logistic Regression and Naive Bayes.

We propose a novel algorithm to learn an FT that best (most accurately) classifies records in a tabular dataset composed of observational tuples, in which values (failures) for each Boolean basic event and Boolean top event in the system are known. We compare it with these existing learning algorithm, in terms of its performance when fitting data and robustness to noisy data.

\vspace{-0.15cm}
\section{Background}
\label{sec:def}
\vspace{-0.2cm}

In this section, we first define the structure of a static FT (consisting of logic gates, and also of intermediate events) and a dataset from which an FT is then inferred.
The formulations below follow definitions from \cite{lift}. 

FTs \cite{VS02} are trees that model how component failures propagate to system failures.
Since subtrees can be shared, FTs are in fact directed acyclic graphs (DAGs) rather than trees. 
Essentially, \emph{intermediate events} in the FT are logical combinations of other intermediate events, with only \emph{basic events} (BE) as the leaves of the tree, and one special intermediate event called the \emph{top event} as root.
Gates model how BE failures lead to system failures. 
Standard fault trees feature two types of gates: AND, and OR.



\vspace{-0.2cm}
\begin{definition}
A \textbf{gate} $G$ is a tuple $(t, \textbf{I}, O)$ such that:
\vspace{-0.2cm}
\begin{itemize}
  \item $t$ is the type of $G$ with $t \in \{And, \ Or \}$.
  \item $\textbf{I}$ is a set of $n \geq 2$ intermediate events $\{i_1, \ldots, i_n\}$ that are inputs of $G$.
  \item $O$ is the intermediate event that is the output of $G$.
\end{itemize}

We denote by $I(G)$ the set of intermediate events in the input of $G$ and by $O(G)$ the intermediate event in the output of $G$.
\end{definition}

\begin{definition}
An \textbf{AND} gate is a gate $(And,\textbf{I},O)$ where output $O$ occurs (i.e. $O$ is True) if and only if every $i \in \textbf{I}$ occurs.
\end{definition}

\begin{definition}
An \textbf{OR} gate is a gate $(Or,\textbf{I},O)$ where output $O$ occurs (i.e. $O$ is True) if and only if at least one $i \in \textbf{I}$ occurs.
\end{definition}

Definition 2 requires that all system components modelled by the events in the input of the \textbf{AND} gate must fail in order for the system modelled by the event in the output to fail. Similarly, Definition 3 requires that one of the system components modelled by the events in the input of the \textbf{OR} gate must fail in order for the system modelled by the event in the output to fail.

\begin{definition}
A \textbf{basic event} $B$ is an event with no input and one intermediate event as output. We denote by $O(B)$ the intermediate event in the output of $B$.
\end{definition}

Sometimes other gates are considered, like the XOR (exclusive OR), the voting gate and the NOT gate \cite{VS02, CCR08}. For the sake of simplicity, we focus on the AND and OR gates; other gates can be treated in a similar fashion.
The root of the tree is called the \emph{top event} (T). The \emph{top event} represents the failure condition of interest, such as the stranding of a train, or the unplanned unavailability of a satellite.
Thus, a FT fails if its \emph{top event} fails.

\begin{definition}
A \textbf{fault tree} \textbf{F} is a tuple $(\textbf{BE},\textbf{IE},T,\textbf{G})$ where:
\vspace{-0.2cm}
\begin{itemize}
  \item $\textbf{BE}$ is the set of basic events; $O(B) \in \textbf{IE},\ \forall B \in \textbf{BE}$. A basic event may be annotated with a probability of occurrence $p$.
  \item $\textbf{IE}$ is the set of intermediate events.
  \item $T$ is the top event, $T \in \textbf{IE}$.\textbf{}
  \item \textbf{G} is the set of gates; $I(G\textbf{}) \subset \textbf{IE} \cup \textbf{BE},\ O(G) \in \textbf{IE},\ \forall G \in \textbf{G}$.
  \item The graph formed by $\textbf{G}$ should be connected and acyclic, with the top event $T$ as unique root.
\end{itemize}
We denote by \textit{IE}(\textbf{F}) the set of intermediate events in $\textbf{F}$ and by \textit{IE}(G) the intermediate event corresponding to gate $G$. 
\end{definition}

\begin{figure}[h]
\vspace*{-0.9cm}
\hspace*{-0.4cm}
\subfloat[Example Fault Tree.\label{fig:exampleft}]{
\begin{tikzpicture}[
    and/.style={and gate US,thick,draw,scale=0.8,fill=black!50,rotate=90,
        anchor=east},
    or/.style={or gate US,thick,draw,scale=0.8,fill=black!20,rotate=90,
        anchor=east},
    be/.style={circle,thick,draw,fill=yellow!90!red,anchor=north,
        minimum width=0.45cm},
    tr/.style={buffer gate US,thick,draw,fill=purple!60,rotate=90,
        anchor=east,minimum width=0.8cm},
    label distance=1mm,
    every label/.style={blue},
    event/.style={rectangle,thick,draw,fill=yellow!20,text width=1.5cm,minimum height=0.35cm,
        text centered,font=\sffamily\bfseries,anchor=north},
    edge from parent/.style={very thick,draw=black!70},
    edge from parent path={(\tikzparentnode.south) -- ++(0,-0.4cm)
            -| (\tikzchildnode.north)},
    level 1/.style={sibling distance=3.5cm,level distance=0.5cm,
            growth parent anchor=south,nodes=event},
    level 2/.style={sibling distance=4.4cm},
    level 3/.style={sibling distance=4cm},
    level 4/.style={sibling distance=2.2cm}
    ]
    \node (g1) [event] {\tiny Lamp Failure (\textbf{T})}
    child{
        child{ node(g2) {\tiny Button Failure}
        child{ 
            child{ node(b0) {\tiny \textbf{OF}} }
            child{ node(b1) {\tiny \textbf{CF}} }
        }
        }
        child{ node(g3) {\tiny Battery Failure}
        child{ 
          child{ node(b3) {\tiny \textbf{LB I}} }
          child{ node(b4) {\tiny \textbf{LB II}} }
        }
        }
    }
        ;
   \node [or]    at (g1.south)    []    {};
   \node [or]    at (g2.south)    []    {};
   \node [and]    at (g3.south)    []    {};
   \node [be]    at (b0.south)    []    {};
   \node [be]    at (b1.south)    []    {};
   \node [be]    at (b3.south)    []    {};
   \node [be]    at (b4.south)    []    {};
\end{tikzpicture}
}\subfloat[Example dataset. \label{table:dataset}]{
\raisebox{1.9cm}{
\rowcolors{2}{gray!25}{white} 
\centering
\begin{scriptsize}
\begin{tabular}{cccccr}
\hline
\textbf{OF} & \textbf{CF} & \textbf{LB I} & \textbf{LB II} & \textbf{T} & count \\
\hline
0 & 0 & 0 & 0 & 0 & 900 \\
0 & 0 & 0 & 1 & 0 & 15 \\
0 & 0 & 1 & 0 & 0 & 5 \\
0 & 0 & 1 & 1 & 1 & 25 \\
0 & 1 & 0 & 1 & 1 & 5 \\
0 & 1 & 1 & 0 & 1 & 5 \\
1 & 0 & 0 & 0 & 1 & 35 \\
1 & 0 & 1 & 0 & 1 & 5 \\
1 & 1 & 0 & 0 & 1 & 3 \\
1 & 1 & 1 & 0 & 1 & 2
\end{tabular}
\end{scriptsize}
}
}
    \centering
    \caption{Example of Fault Tree and learning dataset.}
    \label{fig:exple-ft-dataset}
\end{figure}
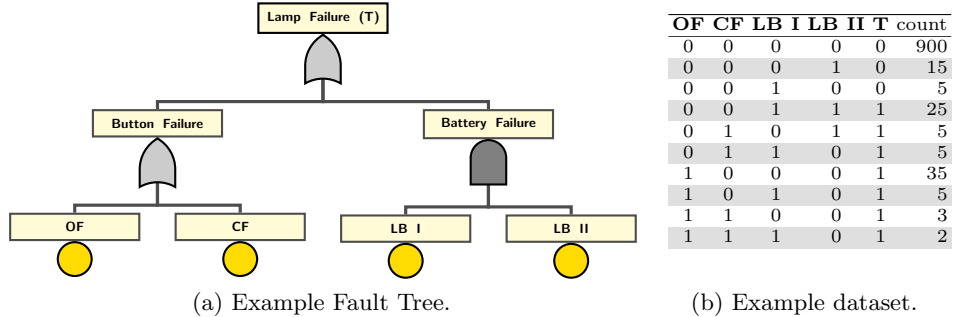

\vspace*{-0.6cm}


\noindent
We now define a data format from which we learn an FT, as a collection of records. Each record is a valuation for the set of BEs (variable that models the state of one basic, indivisible system component), indicating whether a \emph{failure} was observed for that BE. We assume that our dataset is {\em labeled}, i.e., also indicates whether the top event $T$ has failed, yielding the predicted outcome of the FT. 
 
\begin{definition}
A \emph{record $R$} over the set of variables $\textbf{V}$ is a list of length $|\textbf{V}|$ containing tuples $[(V_i,v_i)], 1 \leq i \leq |\textbf{V}|$ where $V_i$ is a variable, $V_i \in V$ and $v_i$ is a Boolean value of $V_i$.
\end{definition}

\begin{definition}
A \emph{dataset $\textbf{D}$} is a set of $|\textbf{D}|$ records, all over the same set of variables $\textbf{V}$. Each variable in $\textbf{V}$ forms a column in $\textbf{D}$, and each record forms a row. When $k$  identical records are present in $\textbf{D}$, a single such record is shown, with a new count column for the value $k$.
\end{definition}

Fig. \ref{fig:exampleft} shows a FT modeling a lamp failure. The top OR-gate shows that a lamp fails if there is either a button failure or a battery failure. A button failure happens if either an operator (\textbf{OF}) or a cable (\textbf{CF}) fails. The AND-gate indicates that battery failure happens if both batteries are low (\textbf{LB I} and \textbf{LB II}).  Table \ref{table:dataset} shows a corresponding dataset.



\vspace{-0.3cm}
\section{Learning Fault Trees with Nature-Inspired Stochastic Optimization}
\label{sec:ea}
\vspace{-0.3cm}

Evolutionary algorithms (EAs) were among the earliest artificial intelligence methods, first envisioned by Alan Turing in 1950~\cite{turing2009computing}.
EAs are heuristics that mimic biological evolution: one starts with an initial population, and iteratively generates new individuals through modification and recombination, where only the best individuals are kept in the next generation -- mimicking survival of the fittest.
EAs are particularly suitable to automatically learn models of some kind, such as trees, graphs or matrix structures, free-form equations, sets and permutations, and synthetic computer programs, etc. Evolutionary algorithms have been very successfully applied in several domains, ranging from antenna designs for spacecrafts~\cite{hornby2006automated}, graph-like network topologies~\cite{BUCUR2014210} and matrix-like robot designs~\cite{cheney2013unshackling}.

The main challenges in devising EAs are (a) formalizing what is a syntactically correct \emph{solution} to the problem (in our case, a well-formed fault tree), and (b) formalizing what makes, semantically, a solution better than another, i.e., writing a \emph{fitness function} which takes any proposed solution and returns a numerical ``goodness'' for that solution. In our case, we want the fault tree to be consistent with the observational data.
Hence the fitness function is the proportion of records correctly classified.
Then, the EA will aim to \emph{maximize} the fitness as close as possible to its optimal value of 100\%.
The EA consists of an iterative optimization process, which maintains a \emph{population} of candidate solutions at each iteration, and randomly mutates and combines (i.e., applies \emph{genetic operators} to) solutions from a population, such that the fitness of the \emph{best solution} per population improves in time. 

\vspace{-0.15cm}
\begin{enumerate}[1.]
    \item \textit{Initialization}: The initial population contains two simple FTs;  all variables in the input dataset $\textbf{D}$ are represented as BEs in these FTs. 
    \item \textit{Mutation and recombination}: Genetic operations are performed on the FTs and generate new FTs.
    \item \textit{Evaluation}: The fitness of each new FT is evaluated.
    \item \textit{Selection}: High-fitness FTs from the new generation replace low-fitness FTs from the previous generation.
    \item \textit{Termination}:
    Steps 2 to 4 are repeated until a given termination criterion is met. This can be if at least one solution in the population exceeds a given fitness bound, or if a given maximum number of iterations is reached. Upon termination, the best solution or solutions in the population are returned.
\end{enumerate}
We describe these steps in more detail below. 




\vspace{-0.15cm}
\subsection{Initialization}
\vspace{-0.05cm}

Our EA takes as input a dataset \textbf{D} and aims at computing an FT with maximal fitness to this dataset. The dataset yields the set of BEs, as well as the top level event $T$.
We start with an initial population consisting of the following two FTs, where all BEs are connected to $T$ via an AND and an OR gate respectively $F_1$ = $(\textbf{BE},\{T\},T,\{(\textit{And},\textbf{BE},T)\})$ and $F_2$ = $(\textbf{BE},\{T\},T,\{(\textit{Or},\textbf{BE},T)\})$, where $\textbf{BE} \cup \{T\} = \textbf{V}$, so all the basic events and the top event must be in the dataset \textbf{D}. These two FTs are the simplest structures including all the observational variables in the data, with an AND gate and an OR gate, respectively, at the top of the FT, and all BEs as inputs of this gate. These two individuals act as a seed population; later populations are larger in size.
The population size is a setting depending on the nature of the FT to learn (namely, the number of BEs). Since the time complexity of the algorithm depends on the population size, increasing the population size may lead to scalability issues. It has also been shown in \cite{chen2012large} that increasing the population size does not always perform as well as expected. In our experiments, we limit the population to hundreds of FTs.

\vspace{-0.25cm}
\subsection{Mutation and recombination}
\vspace{-0.15cm}

\newcommand{\D}{{\textbf{D}}}
\newcommand{\FT}{{\textbf{F}}}
\newcommand{\IE}{{\textbf{IE}}}
\newcommand{\IEit}{{\textit{IE}}}
\newcommand{\ANDgate}{{\textit{And}}}
\newcommand{\ORgate}{{\textit{Or}}}

We define seven stochastic genetic operators (one binary and six unary) which apply to FTs.
For each iteration, each of the genetic operators operates on all individuals in the population with a given probability, to create new individuals. The order in which they are applied is randomized at each iteration.
All our operators are illustrated by Fig. \ref{fig:g-create} to \ref{fig:crossover}.

\vspace{-0.15cm}
\paragraph{G-create.} Given the input FT $\textbf{F} = (\textbf{BE},\textbf{IE},T,\textbf{G})$, create a gate $G \not\in \textbf{G}$, randomly select its nature (AND or OR), then randomly select
a gate $G' \in \textbf{G}$.
Randomly select inputs events $I'$ of $I(G')$ to become inputs of $G$ such that $I(G) = I'$ and $I(G') = I(G') \setminus I'$.
Then, add $O(G)$ to the input events of $G'$ such that $I(G') = I(G') \cup O(G)$.
The new FT is $\textbf{F} = (\textbf{BE},\textbf{IE} \cup O(G),T,\textbf{G}\cup \{G\})$ with $G \not\in \textbf{G}$.
This operator is illustrated in Fig. \ref{fig:g-create}.

\vspace{-0.15cm}
\paragraph{G-mutate.}
Given the input FT $\textbf{F} = (\textbf{BE},\textbf{IE},T,\textbf{G})$, randomly select a gate $G \in \textbf{G}$ and change its nature (AND to OR, or OR to AND).
This operator is illustrated in Fig. \ref{fig:g-mutate}.

\vspace{-0.15cm}
\paragraph{G-delete.}
Given the input FT $\textbf{F} = (\textbf{BE},\textbf{IE},T,\textbf{G})$, randomly select a gate $G \in \textbf{G}$ such that $O(G) \neq T$ and delete it.
We set then, $I(G_p) = \bigcup\limits_{i \in I(G)} IE_{i}$ such that $O(G) \in G_p$.
The new FT is $\textbf{F} = (\textbf{BE},\textbf{IE} \setminus O(G) ,T,\textbf{G}\setminus \{G\})$.
This operator is illustrated in Fig. \ref{fig:g-delete}.

\vspace{-0.15cm}
\paragraph{BE-disconnect.}
Given the input FT $\textbf{F} = (\textbf{BE},\textbf{IE},T,\textbf{G})$ and a randomly chosen basic event $B \in \textbf{BE}$, disconnect $B$ from its intermediate event $G = O(B)$.
The new FT if $\textbf{F} = (\textbf{BE} \setminus \{B\},\textbf{IE},T,\textbf{G})$, where $B \not\in I(G)$.

Note here that a gate $G$ left with 0 or 1 input will not be removed: this is to preserve the solution search space and enable the connection of BEs to this gate. Only the genetic operator \textit{G-delete} can remove such a gate.
This operator is illustrated in Fig. \ref{fig:BE-disconnect}.

\vspace{-0.15cm}
\paragraph{BE-connect.}
Given the input FT $\textbf{F} = (\textbf{BE},\textbf{IE},T,\textbf{G})$ and a basic event $B \not\in \textbf{BE}$ and $B \in \textbf{V} \setminus T$, randomly choose a gate $G \in \textbf{G}$ and connect $B$ to the input of $G$.
The new FT is $\textbf{F} = (\textbf{BE} \cup \{B\},\textbf{IE},T,\textbf{G})$, where $B\in I(G)$. This operator is illustrated in Fig.~\ref{fig:BE-connect}.  Note that this operator is essentially the inverse of {\bf BE-disconnect}.
The relevance of this operator lies in the fact that some FTs in the population may not contain as many BEs as variables $\textbf{V}$ in the dataset. Also, our definition of this operator implies that no BE will be input to 2 different gates. However, the connection of the same BE to 2 different gates can occur within a \textit{crossover} operation.

\vspace{-0.05cm}
\paragraph{BE-swap.}
Given the input FT $\textbf{F} = (\textbf{BE},\textbf{IE},T,\textbf{G})$ and a randomly chosen basic event $B \in \textbf{BE}$ and a randomly chosen gate $G \in \textbf{G} \setminus O(B)$, disconnect $B$ from $O(B)$ and connect $B$ to $G$. This operator is illustrated in Fig. \ref{fig:BE-swap}.

\vspace{-0.05cm}
\paragraph{Crossover.}
The crossover operator takes two FTs as input and swaps at random two of their subtrees. This leads to two new FTs, where the first fault tree contains the selected subtree of the second fault tree and vice versa. More precisely, one selects at random an intermediate event $\IEit_1\in\IE_\textbf{1}$ from the first fault tree $\FT_\textbf{1}$, and one also selects at random an intermediate event $\IEit_2\in\IE_2$ from the section fault tree. Then one replaces in $\FT_\textbf{1}$ the subtree under $\IE_\textbf{1}$ by the subtree under $\IE_\textbf{2}$. Similarly, one replaces in $\FT_\textbf{2}$ the subtree under $\IE_\textbf{2}$ by the subtree under $\IE_\textbf{1}$.
Given two input FTs $\textbf{F}_\textbf{1} = (\textbf{BE}_\textbf{1},\textbf{IE}_\textbf{1},T_1,\textbf{G}_\textbf{1})$
and $\textbf{F}_\textbf{2} = (\textbf{BE}_\textbf{2},\textbf{IE}_\textbf{2},T_2,\textbf{G}_\textbf{2})$,
randomly select an $IE_1 \in \textbf{IE}_\textbf{1}$ and $IE_2 \in \textbf{IE}_\textbf{2}$.
Then, we set $I(O(IE_1)) = I(O(IE_1)) \setminus IE_1 \cup IE_2$ and $I(O(IE_2)) = I(O(IE_2)) \setminus IE_2 \cup IE_1$.
Finally, $O(IE_1) = O(IE_2)$ and vice versa.
This operator is shown in Fig. \ref{fig:crossover}.
Note that in this example, the resulting child in Fig. \ref{fig:crossover}d contains a BE ($BE_3$) connected to multiple gates.

\vspace*{-0.35cm}
\subsection{Evaluation}
\vspace*{-0.15cm}

We define the \emph{fitness} of an FT as the number of records in the dataset for which the value of the top event, given the values of the BEs, is correctly computed.
The count of a record, i.e. the number of appearances, give more weight in the fitness function to records that occur often. In this way, noisy data can be better handled, which often happen in real life applications.

\vspace*{-0.15cm}
\begin{definition} The fitness of a fault tree is its accuracy w.r.t. the dataset $\textbf{D}$ s.t. \vspace{-0.1cm}
$$
f = \frac{\sum\limits_{r \in D} x}{\sum\limits_{r \in D} k} \mbox{ where }
\left\{
  \begin{array}{rr}
    x = k & \mbox{if } V[T] = P[T] \\
    x = 0 & \mbox{otherwise} \\
  \end{array}
\right.
$$
where $P[T]$ stands for the predicted value of the top event given the dataset $\textbf{D}$ for a given FT, $V[T]$ the real value of the top event and $k$ the number of occurrences of the record $r$.

\end{definition}

\vspace*{-0.40cm}
\subsection{Selection}
\vspace*{-0.15cm}

The selection strategy of the best individuals to undergo genetic operations is essential to increase the improvement rate of the fitness of the population.
Commonly used strategies are roulette wheel, stochastic universal sampling, tournament and random selections.
In all our experiments, we use an elitist strategy.
The nature of the individuals motivates this choice: best-fitted FTs are the closest in the population to the optimal solution. They consist of Boolean gates and BEs, which means that a least fitted solution needs more genetic operations to become optimal.
We thus hope that mutating the best FTs will be less costly in terms of iterations of the EA in order to converge towards the right solution.
In Appendix \ref{app:selection-strategies}, we conducted an experiment justifying the relevance of elitism, where we make a comparison between different selection strategies.

\vspace{-0.25cm}
\subsection{Termination}
\vspace{-0.1cm}

The decision of whether to return the best FTs in the actual population or to continue the evolutionary process follows the termination criteria.
In our experiments, we used the standard termination criteria, which are:
\begin{enumerate}
    \item at least one solution in the population achieves an accuracy of 1, which means a perfect fitness to the data.
    \item a maximum number of allowed iterations is reached.
    \item convergence: no improvement of the best FT in the population has been observed for a given number of iterations.
\end{enumerate}

Note that several runs of our EA may return different FTs with the same fitness, especially in terms of structure. Indeed, two FTs with a different structure may be semantically equivalent. We show in Fig. \ref{fig:learnt-fts} how an equivalent FT to a target FT is learnt by our EA. Indeed, FTs, like Boolean formulas, can be factorized to Disjunctive Normal Form (DNF, where a Boolean formula is standardized as a disjunction of conjunctive clauses) or to Conjunctive Normal Form (CNF, where a Boolean formula is standardized as a conjunction of disjunctive clauses). As a result, it may happen that all variables in the dataset do not appear in the FT since they are either not needed or not relevant.
In our experiments, we chose to compute the CNF of the best-fitted FT.
The transformation of an FT into a CNF is based on the following rules: the double negative law, De Morgan's laws, and the distributive law.

\section{Learning of Partial Fault Trees}
\label{sect-partialFT}

A fruitful application of learning fault trees is the learning of partial models, where domain experts partially know the structure of the FT, and other parts need to be inferred from data. In this way expert knowledge 
and data-driven approaches are aggregated.
To accommodate this approach, we propose here a variation of our EA.
We parameterize our EA with such a partially known structure as input to the algorithm.  
The task for the EA is then to evolve \emph{sub}-Fault Trees, given the known \emph{skeleton} of the FT.
The initial population becomes then the partial structures given as input.
Genetic operators are slightly modified to ensure the given skeletons to remain unmodified in each mutated FT.
We gather at any moment of the evolutionary process a population composed of FTs containing the allowed skeleton.
In this section, we detail the initialization and the genetic operators to take into account the expert knowledge provided as a \textit{skeleton} FT. 

\subsection{Initialization}
In the same way as the procedure described in Sect. \ref{sec:ea}, our evolutionary algorithm takes as input a dataset \textbf{D}, as well a known FT-\textit{skeleton} $F_o$. We start then with an initial population consisting of this FT-\textit{skeleton}:
$$
F_o = (\mathbf{BE_o},\mathbf{IE_o},T_o,\mathbf{G_o})
$$
where $\mathbf{BE_o}$ are BEs contained in the skeleton,  $\mathbf{IE_o}$ are the intermediate events of the skeleton,  $T_o$ is the top event of the skeleton and  $\mathbf{G_o}$ is the set of gates contained in the skeleton.

This is this structure given as input that will remain in all mutated FTs all along the evolutionary process.

\subsection{Mutation and recombination}
In order to preserve the FT-\textit{skeleton} during mutation and recombination operations, we have to adapt the following genetic operators.

\paragraph{G-create-o.}
Given the input FT $\textbf{F} = (\textbf{BE},\textbf{IE},T,\textbf{G})$, create a gate $G \not\in \textbf{G}$, randomly select its nature (AND or OR), then randomly select
a gate $G' \in \textbf{G}$.
Randomly select inputs events $I'$ of $I(G') \setminus \mathbf{IE_o}$ to become inputs of $G$ such that $I(G) = I'$ and $I(G') = I(G') \setminus I'$.
Then, add $O(G)$ to the input events of $G'$ such that $I(G') = I(G') \cup O(G)$.
The new FT is $\textbf{F} = (\textbf{BE},\textbf{IE} \cup O(G),T,\textbf{G}\cup \{G\})$ with $G \not\in \textbf{G}$.
In that way, we allow the creation of a gate such that its input events are not in the gates of the skeleton, that is, the skeleton remains unchanged. 

\paragraph{G-mutate-o.}
Given the input FT $\textbf{F} = (\textbf{BE},\textbf{IE},T,\textbf{G})$, randomly select a gate $G \in \mathbf{G} \setminus \mathbf{G_o} $ and change its nature (AND to OR, or OR to AND).

\paragraph{G-delete-o.}
Given the input FT $\textbf{F} = (\textbf{BE},\textbf{IE},T,\textbf{G})$, randomly select a gate $G \in \textbf{G} \setminus \mathbf{G_o}$ such that $O(G) \neq T$ and delete it.
We set then, $I(G_p) = \bigcup\limits_{i \in I(G)} IE_{i}$ such that $O(G) \in G_p$.

\paragraph{BE-disconnect-o.}
Given the input FT $\textbf{F} = (\textbf{BE},\textbf{IE},T,\textbf{G})$ and a randomly chosen basic event $B \in \mathbf{BE} \setminus \mathbf{BE_o}$, disconnect $B$ from its intermediate event $G = O(B)$.
The new FT if $\textbf{F} = (\textbf{BE} \setminus \{B\},\textbf{IE},T,\textbf{G})$, where $B \not\in I(G)$.

\paragraph{BE-swap-o.}
Given the input FT $\textbf{F} = (\textbf{BE},\textbf{IE},T,\textbf{G})$ and a randomly chosen basic event $B \in \mathbf{BE} \setminus \mathbf{BE_o}$ and a randomly chosen gate $G \in \textbf{G} \setminus O(B)$, disconnect $B$ from $O(B)$ and connect $B$ to $G$. This operator is illustrated in Fig. \ref{fig:BE-swap}.

\paragraph{Crossover-o.}
Given two input FTs $\textbf{F}_\textbf{1} = (\textbf{BE}_\textbf{1},\textbf{IE}_\textbf{1},T_1,\textbf{G}_\textbf{1})$
and $\textbf{F}_\textbf{2} = (\textbf{BE}_\textbf{2},\textbf{IE}_\textbf{2}$ $,T_2,\textbf{G}_\textbf{2})$,
randomly select an $IE_1 \in \mathbf{IE}_\mathbf{1} \setminus \mathbf{IE_o}$ and $IE_2 \in \mathbf{IE}_\mathbf{2}  \setminus \mathbf{IE_o}$.
Then, we set $I(O(IE_1)) = I(O(IE_1)) \setminus IE_1 \cup IE_2$ and $I(O(IE_2)) = I(O(IE_2)) \setminus IE_2 \cup IE_1$.
Finally, $O(IE_1) = O(IE_2)$ and vice versa.

Note that the scenario where expert guidance is used to lead to faster convergence of the FT learning algorithm is realistic: indeed, this variation is helpful to refine existing FTs, or checking if a handmade model is accurate given real-world measurements.

\section{Experimental Evaluation}
\label{sec:expe}

We have evaluated the efficiency and effectiveness of our EA method using a large number of cases. We compared our methods with six other learning techniques: five approaches from the literature, and the variant of our own EA technique for learning partial fault trees. For these methods, we investigated both the accuracy and as well as runtime. 
Our comparisons were performed for a set of synthetic cases (Sections~\ref{sect-synthetic} and \ref{sect-synthetic2}), as well as for industrial benchmarks (Section~\ref{sect-industrial} and \ref{sect-ftbenchmark}).

\subsection{Experimental set up}
The first three methods in our evaluation are:
(1) Support Vector Machine (abbreviated svm in the figures),
(2) Logistic Regression (abbreviated log)
and (3) Naive Bayes Classifier (nba).
These methods are Boolean classifiers that, given the values of the BEs, predict the value of the top event $T$.
Being classifiers, methods (1) - (3) do not yield FT models, only a prediction for the value of $T$.

Then, we have used three methods that do learn fault tree models: 
(4) We have compared our results to the well-known 
C4.5 algorithm for learning decision trees
(abbreviated c45). 
Decision trees can be transformed to FTs, by first computing in the decision tree the conjunction of all paths leading to failure leaves, and then simplifying the conjunction to CNF.
(5) We have also compared to the earlier LIFT
approach, which re returns an FT.
(6) Finally, we used the variation of the EA for learning partial fault trees (ea-p). Here, we assumed that the two upper layers of the fault trees were fixed. 

To compare these methods, the observational data was divided into two sets: one training set, used as input to the EA (with an average of 2/3 of all possible observations), and a test set containing all observational variables (complete boolean table), used to evaluate the solution returned by our algorithms.
The parameters of the EA were set as follows: we used a population size of 100. As termination criteria, we used either a maximum number of 100 iterations or an observed convergence (i.e. no improvement of the best individual's fitness) over 10 iterations or an FT with fitness 1 (optimal solution) in the population. Each genetic operator was applied with probability  $0.9$ in order to increase the mutation rate of the population. The selection and replacement strategy were elitists, to systematically replace least-fitted individuals in the old population by the best individuals in the union of the old population and the set of newly generated individuals.
Finally, to homogenize several runs of the EA, the conjunctive normal form (CNF) of the best FT was returned in the termination step.
Note that our Python implementations and dataset are available\footnote{\href{https://gitlab.science.ru.nl/alinard/learning-ft}{https://gitlab.science.ru.nl/alinard/learning-ft}}, and that we used state-of-the-art implementations of the scikit-learn library for techniques (1)-(5).

\subsection{Synthetic Dataset: accuracy and runtime}\label{sect-synthetic}

\begin{figure}[h]
\vspace{-0.6cm}
\hspace*{-0.4cm}
\subfloat[Accuracy of the learnt FTs.\label{fig:results-accuracy}]{
    \includegraphics[width=0.51\linewidth]{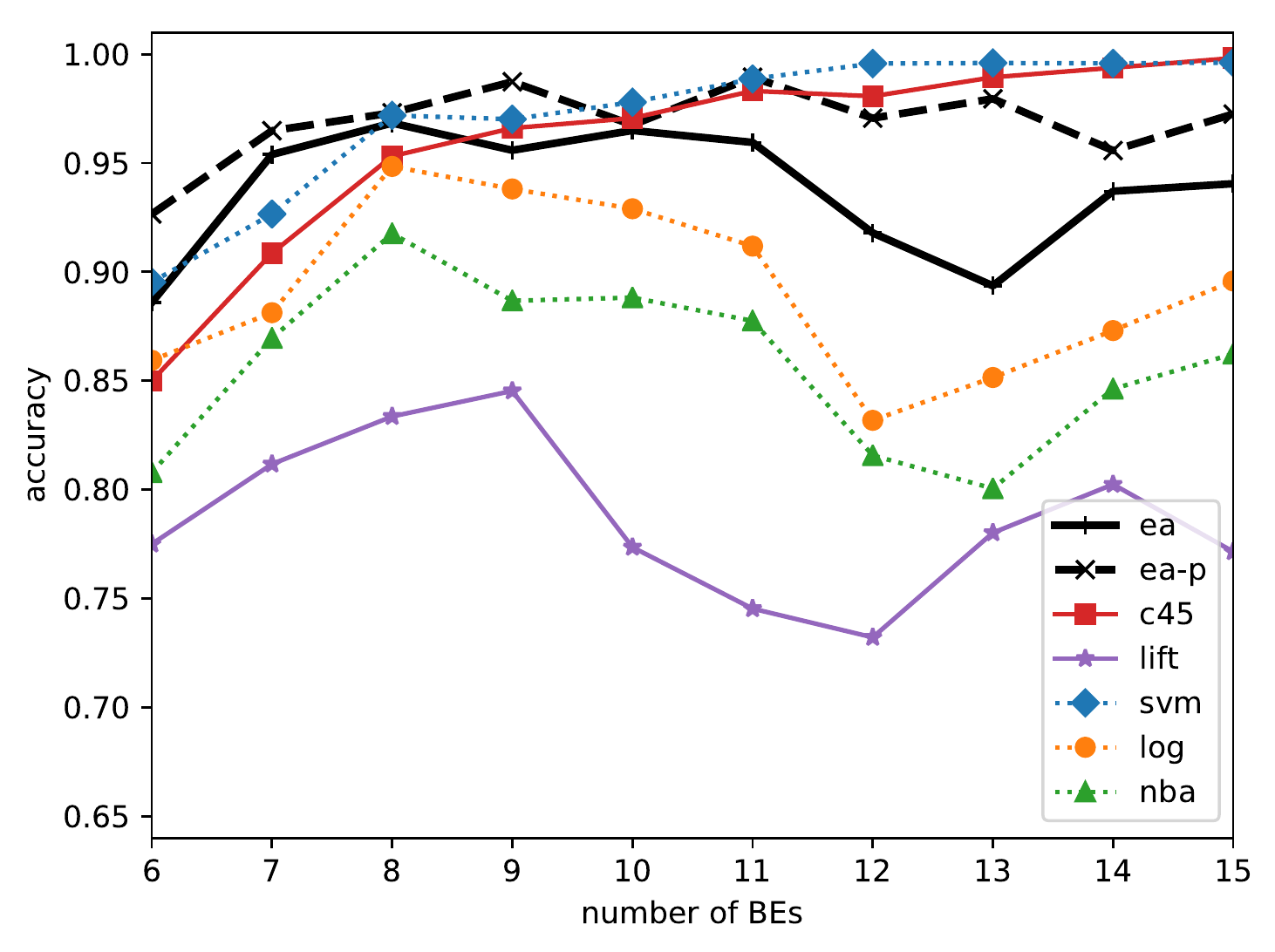}
}\subfloat[Runtime of different algorithms. \label{fig:results-accuracy-numbes}]{
    \includegraphics[width=0.51\linewidth]{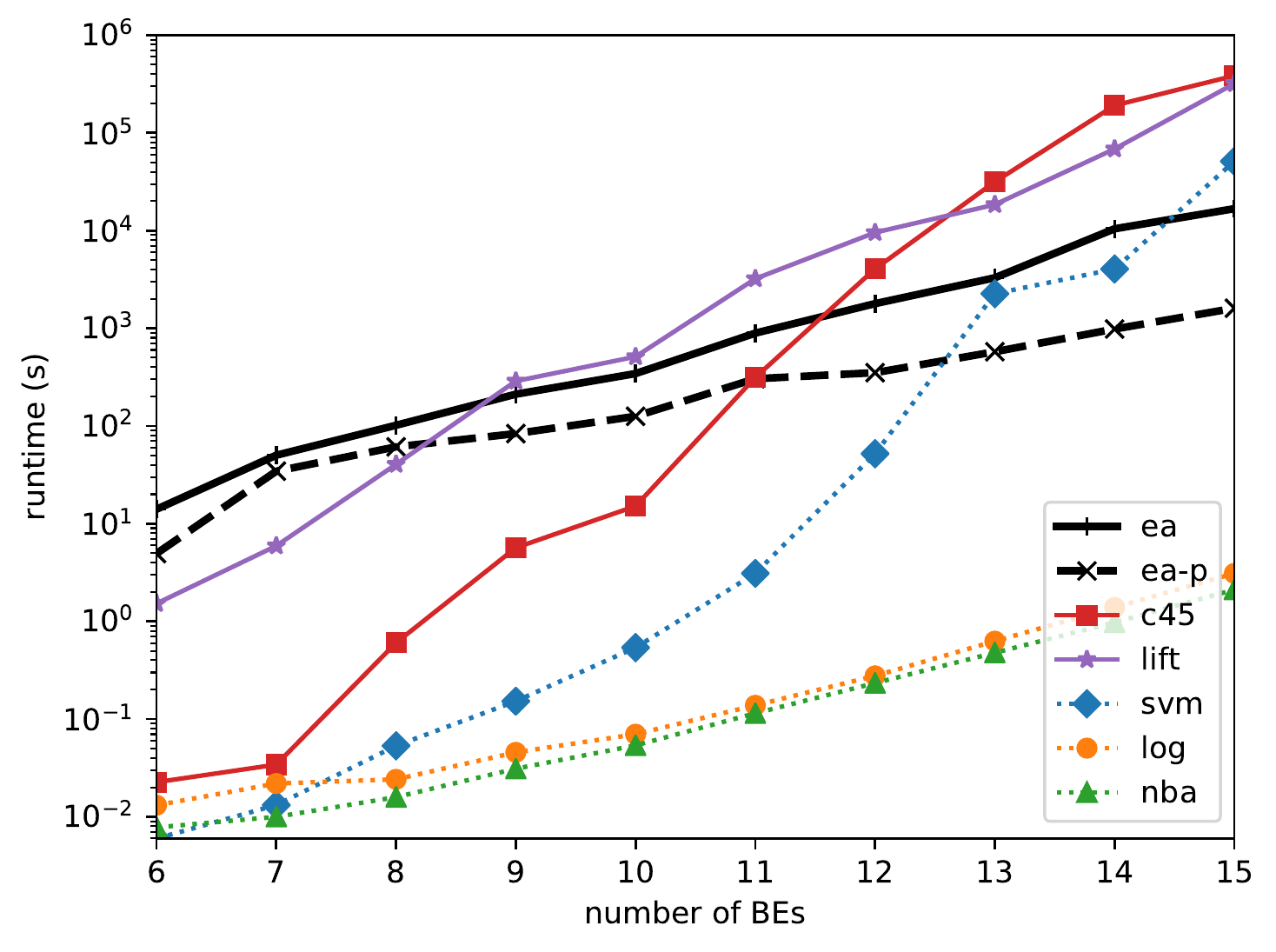}
}
    \centering
    \caption{Comparison of different learning algorithms.}
    \label{fig:results-synthetic}
    \vspace*{-0.3cm}
\end{figure}

We have first used a large synthetic case. We considered 100 randomly generated fault trees with 6 to 15 BEs, and for each FT, a randomly generated data set, with 200 to 230k records. 
Figures \ref{fig:results-accuracy} and \ref{fig:results-accuracy-numbes} present respectively the average accuracy and the average runtime, both as functions of the number of BEs.
Fig.~\ref{fig:results-accuracy} shows that the svm and c45 methods have the highest accuracy. However, the svm method only provides a classifier, not a fault tree. Further, the c45 method does not perform well in terms of runtime, see Fig~\ref{fig:results-accuracy-numbes}. Our methods EA and EA-p perform reasonably well in terms of accuracy, as well as in term of run time. Finally, the log and nba method are fast but provide low accuracy.
We also see that LIFT obtains less good results. An exponential complexity can explain this, and the fact LIFT requires data about intermediate events.

We can also see that the more the FTs contain BEs, the more it is complicated to gather a solution with perfect fitness w.r.t. the training set. This is due to the significant number of iterations needed to converge to an optimal solution when dealing with a large number of BEs.
However, we can see that expert knowledge is extremely beneficial in the case of ea-p, where the skeleton of the FT is given. It enables us to learn more accurate FTs (accuracy $> 95$\%) and faster (up to 10 times faster than the baseline EA).

\vspace{-0.2cm}
\subsection{Synthetic Dataset: other statistics}
\label{sect-synthetic2}

\vspace{-1cm}
\begin{figure}[!h]
\hspace*{-0.45cm}
\subfloat[Effect of noise on the learned FTs.\label{fig:results-noise}]{
    \includegraphics[width=0.52\linewidth]{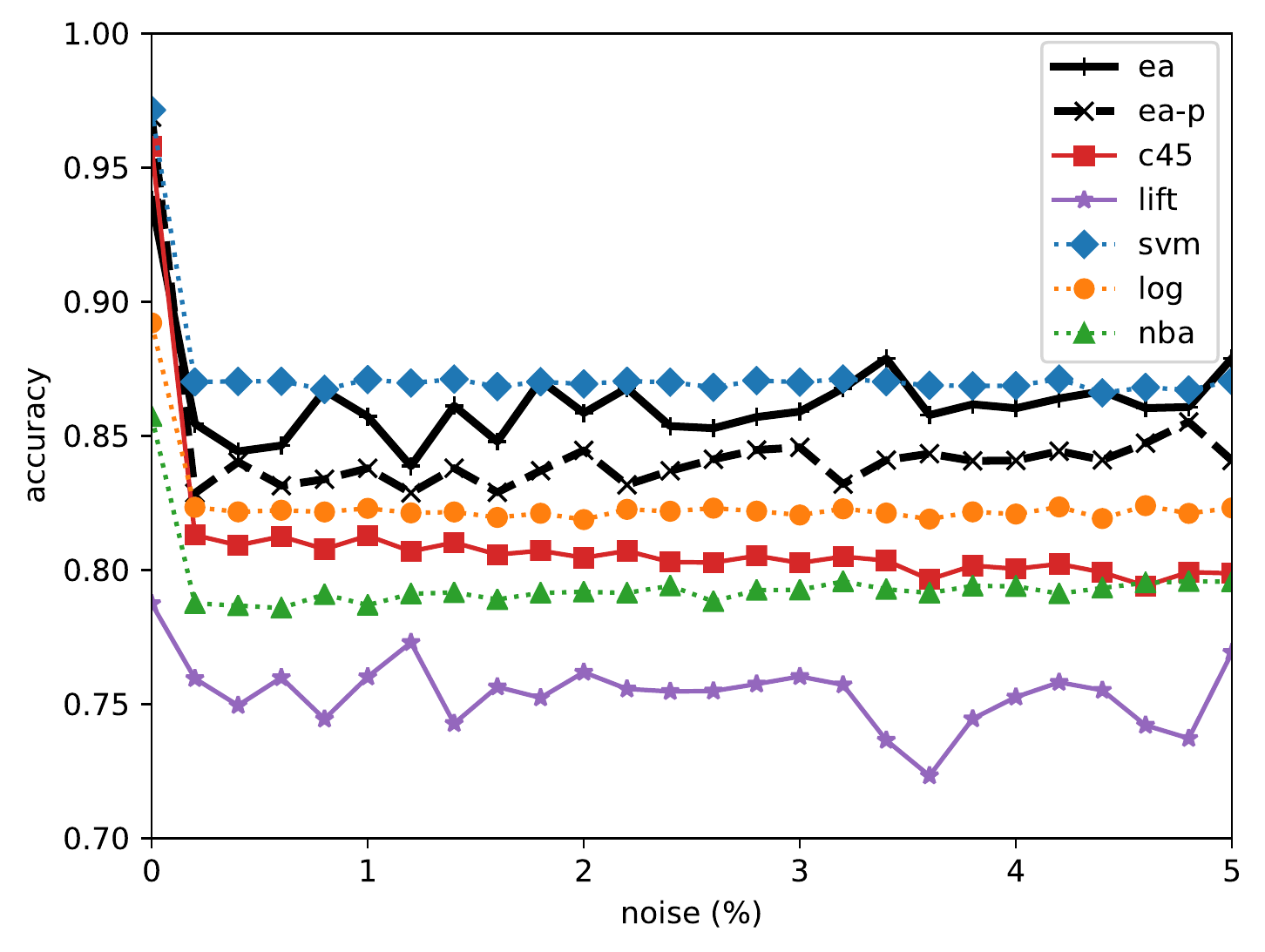}
}
\subfloat[Number of successful GOs per type.\label{fig:results-usedgates-fitness}]{
    \includegraphics[width=0.508\linewidth]{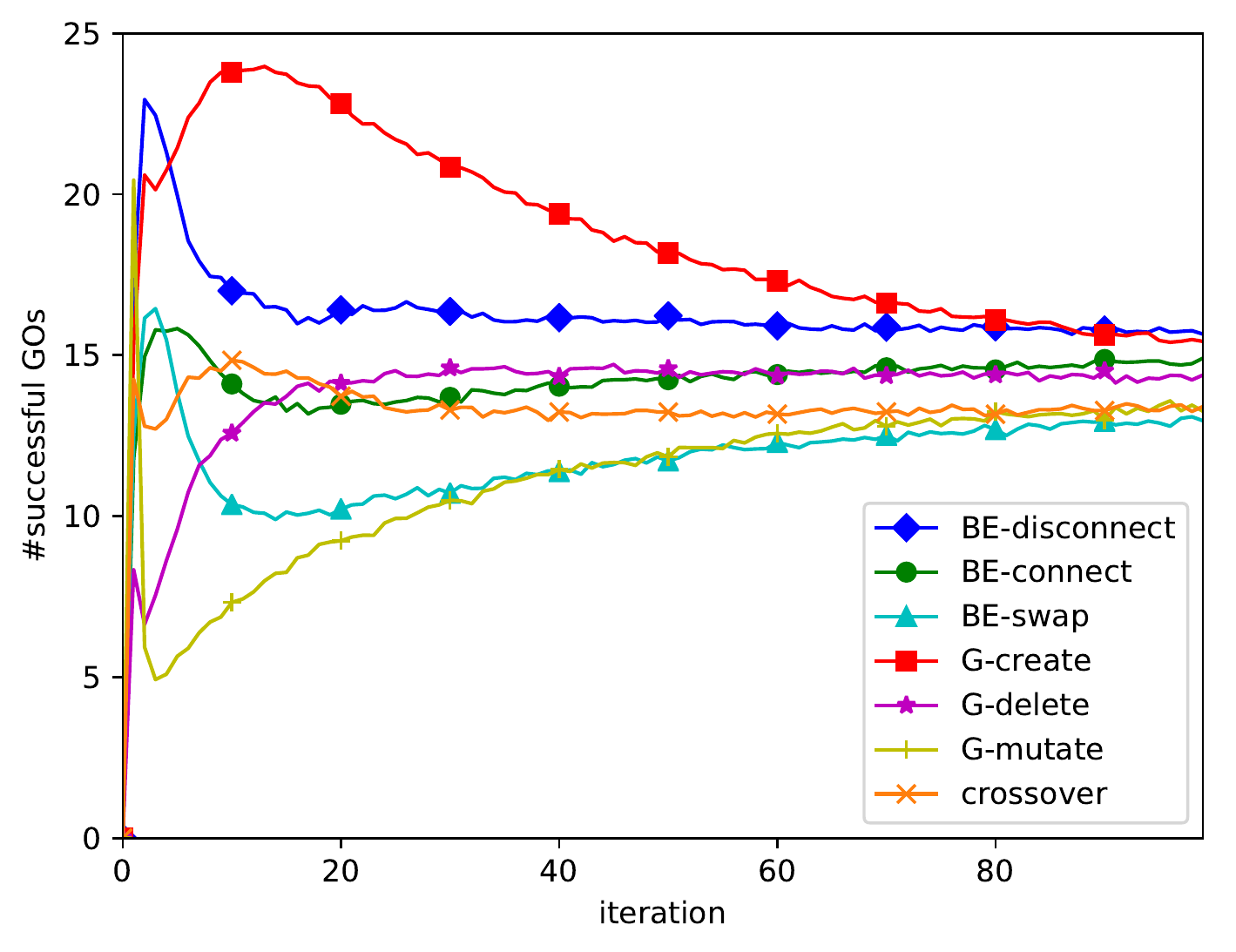}
}
\caption{Statistics on genetic algorithm.}
\end{figure}
\vspace{-0.3cm}

We also carried out an experiment where noise is added in the dataset, in order to test the robustness of the different algorithms. 
The noise varies from 0 to 5\% of noisy records.
We call a \textit{noisy} record one where the value of at least one variable has been changed, i.e. measured incorrectly in real life.
In the results shown in Fig. \ref{fig:results-noise}, we see that our methods are relatively robust against noise compared to other methods. However, we see that the accuracy of the learned FTs drops whenever noise is present in the dataset.

Further, we also investigated which genetic operations were successful,
as a function of the number of iterations, shown in Fig. \ref{fig:results-usedgates-fitness}). The latter is computed by looking at, for each iteration, the number of individuals issued from the same genetic operator who survived in the next generation, i.e. whose fitness was good enough to be kept in the population.
We see that the success of most operations depends on the stage of the EA: this is the case for \textit{BE-disconnect}, and \textit{G-create}, which provide satisfying new individuals during the first iterations of the algorithm. An explanation is that \textit{G-create} will increase the size of the FT, i.e. its complexity. Then, the search space of the solution is increased.  In opposition to these gates, \textit{G-mutate} seems to be a less good operator since the number of individuals issued from it tends not to survive in the population. This is mainly due to the change of semantics this operator implies: indeed, when the depth of the mutated gate is small (i.e. close to the top event), the meaning of the resulting FT may drastically change. Hence a small number of successful operations of this type.

\vspace{-0.1cm}
\subsection{Case Study with Industrial Dataset}
\label{sect-industrial}
\vspace{-0.1cm}

We present here an industrial case study based on the dataset from \cite{isola-oce}. We consider here a component called the nozzle.
The system containing the nozzles records large amounts of data about the state of the components over time, among them the failing of nozzles and nozzle-related factors. 
The dataset is composed of 9,000 records, 8 basic events being nozzles-related factors and a Boolean top event, standing for nozzle failure.
We ran our genetic algorithm 10 times, with a maximum allowed iterations of 100 (convergence criterion of 10), and the fitness of the best FT learnt was of $0.997$ (split ratio for train/test set of 80/20). The resulting FT, shown in Fig.~\ref{fig:ft-nozzle}, has been validated by domain experts.
Even in a practical context, multiple runs of the EA may return different (possibly equivalent) FTs, with different structures. Depending on the applications, expert knowledge can figure out whether one or the other returned FT is the most relevant to the case study. To help the selection process, one can place additional constraints on the FT, such as the number of children.

\vspace{-0.1cm}
\subsection{Fault Tree Benchmark}
\label{sect-ftbenchmark}
\vspace{-0.1cm}

\begin{figure}[t]
\hspace*{-0.4cm}
\subfloat[Results of FT Benchmark (accuracy). \label{fig:results-benchmark-accuracy}]{
    \includegraphics[width=0.51\linewidth]{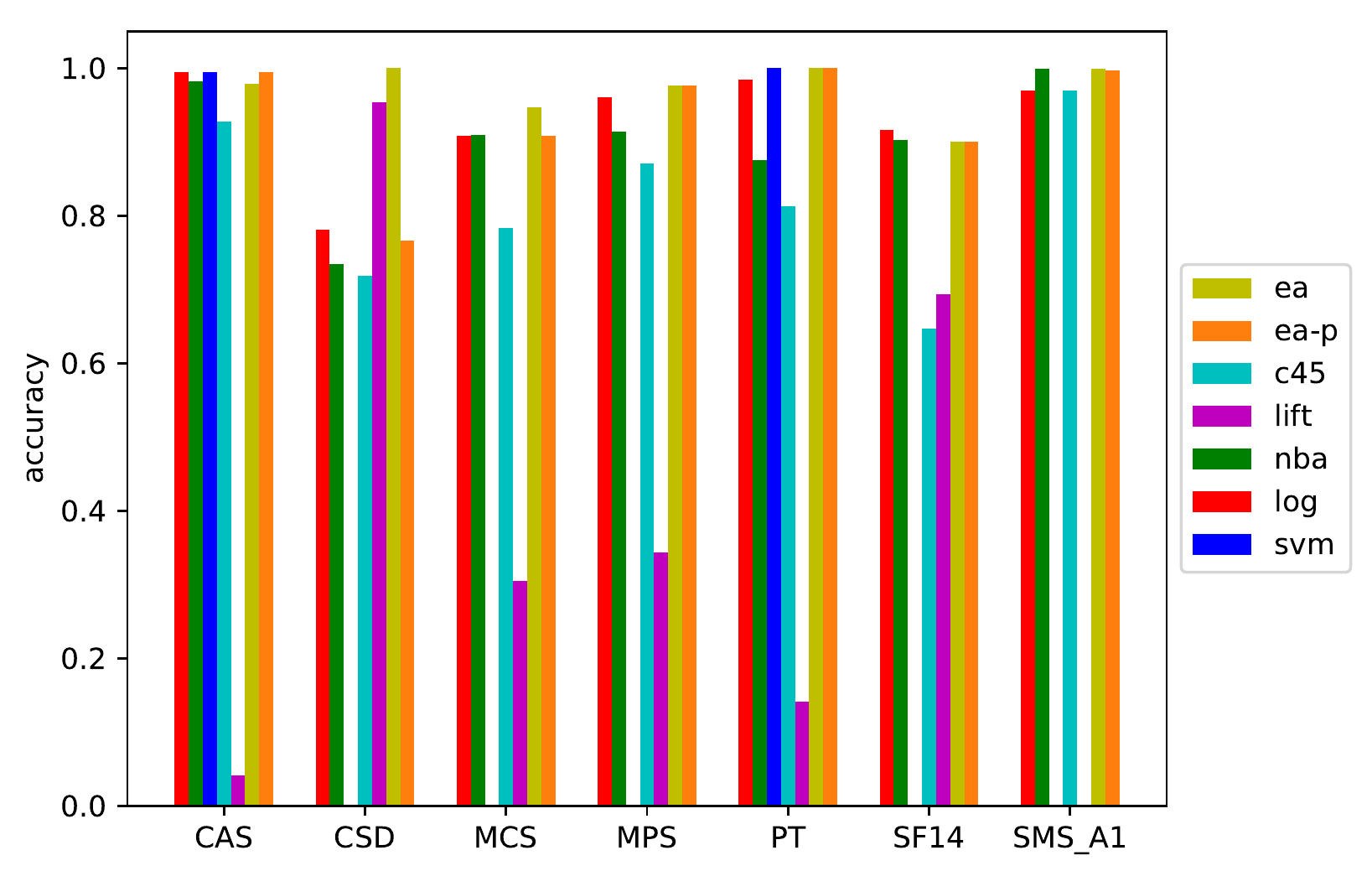}
}\subfloat[Results of FT Benchmark (runtime).\label{fig:results-benchmark-runtime}]{
    \includegraphics[width=0.51\linewidth]{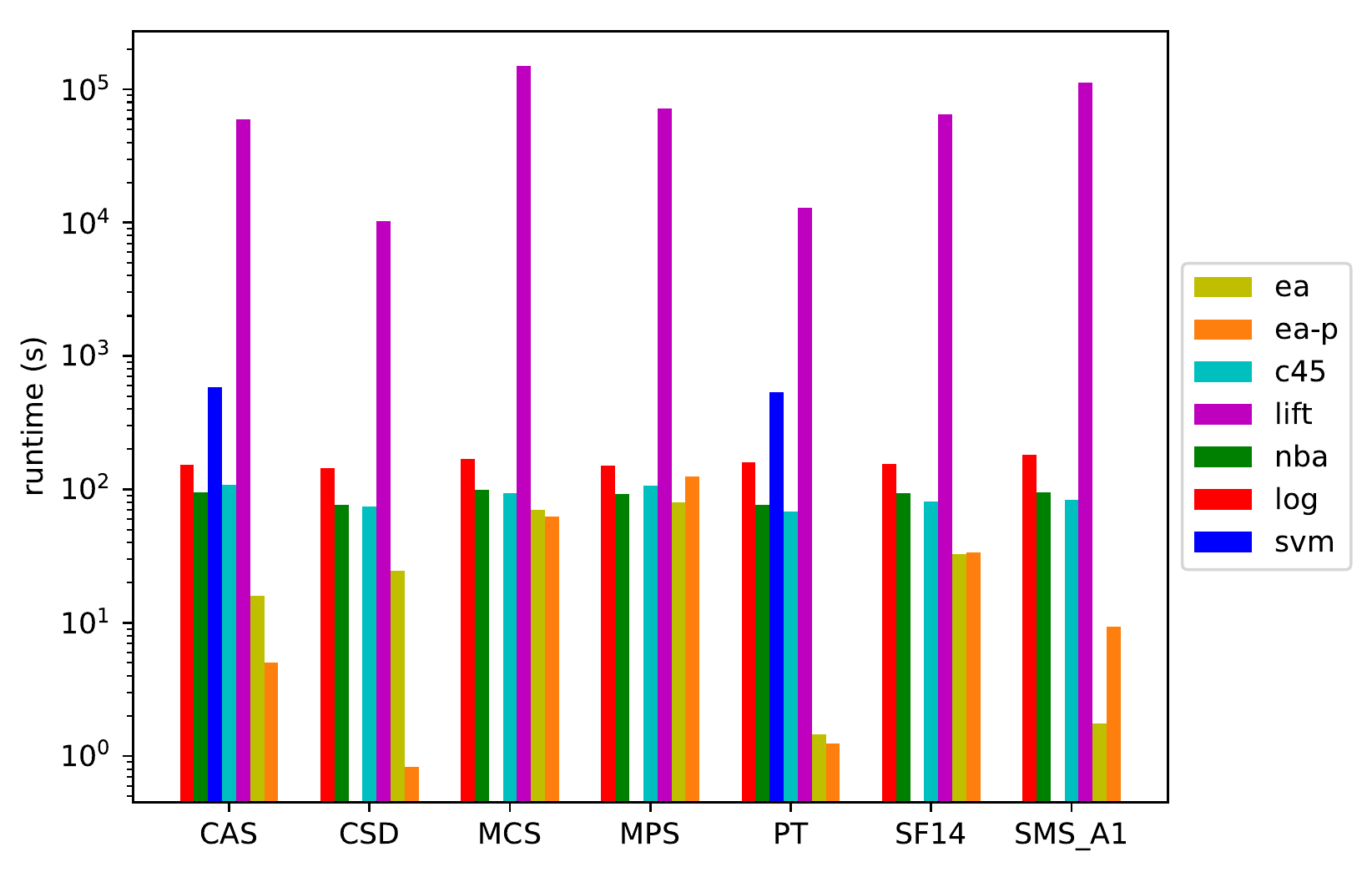}
}
    \caption{Results of Fault Tree Benchmark.}
    \vspace{-0.3cm}
\end{figure}

We present here the results we obtained for a set of publicly available benchmark suite\footnote{https://dftbenchmarks.utwente.nl/}, consisting of industrial fault trees from the literature, containing from 6 to 14 BEs and 4 to 10 gates. The FTs used are Cardiac Assist System (CAS), Container Seal Design Example (CSD), Multiprocessor Computing System (MCS), Monopropellant Propulsion System (MPS), Pressure Tank (PT), Sensor Filter Network (SF14) and Spread Mooring System (SMS\_{A1}). 
Whereas the fault tree models were given in the literature, no data sets were available. Therefore, we have randomly generated these data sets, containing 10M records per case, in order to cope with low failure rates.
Since the benchmark does provide failure probabilities per BEs, we have used those probabilities: If $p_e$ is the failure probability of BE $e$ in the benchmark, then we set, in each data record, $R[e]=1$ with probability $p_e$.
Fig. \ref{fig:results-benchmark-accuracy} and \ref{fig:results-benchmark-runtime} present the accuracy and runtime,  respectively.
Missing bars stand for experiments for which no result could have been obtained within 1 week of running time.
We can see that for all case studies, our method is either the most or the second most efficient. 
We also see that in all cases, our method is among the most accurate methods. 



\vspace{-0.2cm}
\section{Discussion}
\label{sec:discussion}
\vspace{-0.2cm}

\paragraph{Extensions.} The definition of gates and genetic operators can be extended. We show how to deal with \textbf{K/N} gates, which are gates of type $(k/N,\textbf{I},O)$ where output $O$ occurs (i.e. $O$ is True) if at least $k$ input events $i \in \textbf{I}$ occurs, with $|\textbf{I}| = N$. The cardinality of a $k/N$ gate is said to be the number $k$.
Note that this gate can be replaced by the OR of all sets of $k$ inputs, but the use of $k/N$ gates is much more compact for the representation of a FT. We can then define new genetic operators, such as \textbf{k-n-change} where, given an input FT $\textbf{F} = (\textbf{BE},\textbf{IE},T,\textbf{G})$, randomly select a $k/N$ gate $G \in \textbf{G}$ and change its cardinality, such that $k \in [1,N-1]$.
We also extend the mutate gate operator of \textbf{G-mutate}, as follows: Given an input FT $\textbf{F} = (\textbf{BE},\textbf{IE},T,\textbf{G})$, randomly select a gate $G \in \textbf{G}$ and change its nature (AND to OR or $k/N$ ; OR to AND or $k/N$ ; $k/N$ to OR or AND). 
Similarly, we can redefine the create gate operator \textbf{C-create} such that the randomly selected nature of the new gate is chosen among AND, OR or $k/N$.

\noindent
Our formalism can also handle \textbf{NOT} gates so that the FTs can become non-monotonic.

\paragraph{Limitations.} While we can accurately learn small fault trees, the main limitation of our method at the moment is scalability. While other techniques, especially naive Bayesian classifiers, score well,
techniques that learn models experience slower performance. Therefore, a solution may be to combine both methods. We can also use better heuristics on which GO to deploy, and with what parameters. Such ideas were also the key to the success of EAs in other application domains.
The result obtained by the EA does not ensure a perfect fitness of the FT with regards to the data. This is the case when a maximal number of iteration has been reached, and the best FT in the population returned. Hence the near-optimality of our algorithm.
In addition, multiple runs of the EA may return different (possibly equivalent) FTs, with different structures.
We leave for further work the discovery of which of the returned FT is right, based on the data, figuring out causal relationships between variables using Mantel-Haenszel Partial Association score \cite{birch1964detection}.
Another limitation lies in the growing size of the FTs after iterations: this may lead to overgrown FTs. However, we think that the best fit individuals may be compacted when returned: indeed, some gates in the FTs may contain none of only one input, and some factorization can be applied. We thus recommend performing FT reduction on the returned FTs, such as the calculation of CNFs or DNFs. 
A first alternative would be to reduce to CNF or DNF the FTs in the population at each iteration of the EA. However, this would drastically reduce the search space of the EA (e.g. by making fewer mutations/recombinations possible, hence leading to a lower fitness) and go against genetic programming principles.
A second alternative would be to take into account the size of the solutions as a second fitness function for the selection step. The implementation of such a multi-objective EA \cite{nsga2} is left for further work.

\noindent
In all cases, learning FTs from an already known \textit{skeleton} FT may suffer less from overgrown FTs after several iterations: the \textit{skeleton} may give, indeed, already enough information on the structure of the FT. Thus, it helps the algorithm to converge faster to a solution.

\vspace{-0.2cm}
\section{Conclusion and Future Work}
\label{sec:ccl}
\vspace{-0.2cm}

We presented an evolutionary algorithm for the automated generation of FTs from Boolean observational data. We defined a set of genetic operators specific to the formalism of FTs.
Our results show the robustness and scalability of our algorithm.
Our future research will focus on the learning of dynamic FTs, and especially trying to learn their specific gates such as PAND, FDEP and SPARE gates.
We will also further look into Bayesian Inference and translating rules from Bayesian Networks to FTs.
We also hope to take into account different failure modes of components thanks to INHIBIT gates.
Finally, since there are many possible (i.e. logically equivalent) alternatives to an FT, we would like to investigate further what are the features of a good FT. In other words, we think that we need to characterize how much better a particular FT structure is compared to another.

\vspace{-0.1cm}
 \bibliographystyle{splncs04}
 \bibliography{biblio}

\newpage
\appendix

\section{Illustrated Genetic Operators}
\label{app:illustrated-gos}

\vspace{-1cm}
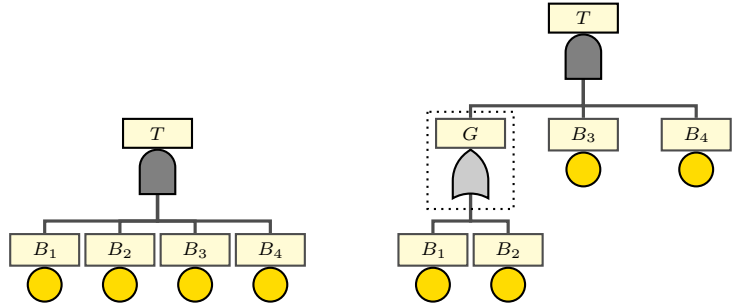
\begin{figure}[!ht]
\subfloat{
\centering
\begin{tikzpicture}[
    and/.style={and gate US,thick,draw,scale=0.8,fill=black!50,rotate=90,
		anchor=east},
    or/.style={or gate US,thick,draw,scale=0.8,fill=black!20,rotate=90,
		anchor=east},
    be/.style={circle,thick,draw,fill=yellow!90!red,anchor=north,
		minimum width=0.45cm},
    tr/.style={buffer gate US,thick,draw,fill=purple!60,rotate=90,
		anchor=east,minimum width=0.8cm},
    label distance=1mm,
    every label/.style={blue},
    event/.style={rectangle,thick,draw,fill=yellow!20,text width=0.7cm,
		text centered,font=\sffamily,anchor=north},
    edge from parent/.style={very thick,draw=black!70},
    edge from parent path={(\tikzparentnode.south) -- ++(0,-0.4cm)
			-| (\tikzchildnode.north)},
    level 1/.style={sibling distance=2cm,level distance=0.56cm,
			growth parent anchor=south,nodes=event},
    level 2/.style={sibling distance=1cm},
    level 3/.style={sibling distance=1cm},
    level 4/.style={sibling distance=1cm}
    ]
    \node (g1) [event] {\scriptsize $T$}
    child{
        	child{ node(b1) {\scriptsize $B_1$} }
            child{ node(b2) {\scriptsize $B_2$} }
        child{ node(b3) {\scriptsize $B_3$} }
        child{ node(b4) {\scriptsize $B_4$} }
    }
        ;
   \node [and]	at (g1.south)	[]	{};
   \node [be]	at (b1.south)	[]	{};
   \node [be]	at (b2.south)	[]	{};
   \node [be]	at (b3.south)	[]	{};
   \node [be]	at (b4.south)	[]	{};
\end{tikzpicture}
}\subfloat{\hspace{1cm}
\centering
\begin{tikzpicture}[
    and/.style={and gate US,thick,draw,scale=0.8,fill=black!50,rotate=90,
		anchor=east},
    or/.style={or gate US,thick,draw,scale=0.8,fill=black!20,rotate=90,
		anchor=east},
    be/.style={circle,thick,draw,fill=yellow!90!red,anchor=north,
		minimum width=0.45cm},
    tr/.style={buffer gate US,thick,draw,fill=purple!60,rotate=90,
		anchor=east,minimum width=0.8cm},
    label distance=1mm,
    every label/.style={blue},
    event/.style={rectangle,thick,draw,fill=yellow!20,text width=0.7cm,
		text centered,font=\sffamily,anchor=north},
    edge from parent/.style={very thick,draw=black!70},
    edge from parent path={(\tikzparentnode.south) -- ++(0,-0.4cm)
			-| (\tikzchildnode.north)},
    level 1/.style={sibling distance=3cm,level distance=0.56cm,
			growth parent anchor=south,nodes=event},
    level 2/.style={sibling distance=1.5cm},
    level 3/.style={sibling distance=1.5cm},
    level 4/.style={sibling distance=1cm}
    ]
    \node (g1) [event] {\scriptsize $T$}
    child{
        child{ node(g2) {\scriptsize $G$}
        child{ 
        	child{ node(b1) {\scriptsize $B_1$} }
            child{ node(b2) {\scriptsize $B_2$} }
        }
        }
        child{ node(b3) {\scriptsize $B_3$} }
        child{ node(b4) {\scriptsize $B_4$} }
    }
        ;
   \node [and]	at (g1.south)	[]	{};
   \node [or]	at (g2.south)	[]	{};
   \node [be]	at (b1.south)	[]	{};
   \node [be]	at (b2.south)	[]	{};
   \node [be]	at (b3.south)	[]	{};
   \node [be]	at (b4.south)	[]	{};
  \draw[thick,dotted]     ($(g2.north west)+(-0.1,0.08)$) rectangle ($(g2.south east)+(0.1,-0.8)$);
\end{tikzpicture}
}
\centering
\caption{Example of \textit{G-create}.}
\label{fig:g-create}
\end{figure}

\begin{figure}[!ht]
\subfloat{
\begin{tikzpicture}[
    and/.style={and gate US,thick,draw,scale=0.8,fill=black!50,rotate=90,
		anchor=east},
    or/.style={or gate US,thick,draw,scale=0.8,fill=black!20,rotate=90,
		anchor=east},
    be/.style={circle,thick,draw,fill=yellow!90!red,anchor=north,
		minimum width=0.45cm},
    tr/.style={buffer gate US,thick,draw,fill=purple!60,rotate=90,
		anchor=east,minimum width=0.8cm},
    label distance=1mm,
    every label/.style={blue},
    event/.style={rectangle,thick,draw,fill=yellow!20,text width=0.7cm,
		text centered,font=\sffamily,anchor=north},
    edge from parent/.style={very thick,draw=black!70},
    edge from parent path={(\tikzparentnode.south) -- ++(0,-0.4cm)
			-| (\tikzchildnode.north)},
    level 1/.style={sibling distance=3cm,level distance=0.56cm,
			growth parent anchor=south,nodes=event},
    level 2/.style={sibling distance=2cm},
    level 3/.style={sibling distance=1.5cm},
    level 4/.style={sibling distance=1cm}
    ]
    \node (g1) [event] {\scriptsize $T$}
    child{
        child{ node(g2) {\scriptsize $G_1$}
        child{ 
        	child{ node(b1) {\scriptsize $B_1$} }
            child{ node(b2) {\scriptsize $B_2$} }
        }
        }
        child{ node(g3) {\scriptsize $G_2$}
        child{ 
          child{ node(b3) {\scriptsize $B_3$} }
          child{ node(b4) {\scriptsize \scriptsize $B_4$} }
        }
        }
    }
        ;
   \node [and]	at (g1.south)	[]	{};
   \node [and]	at (g2.south)	[]	{};
   \node [or]	at (g3.south)	[]	{};
   \node [be]	at (b1.south)	[]	{};
   \node [be]	at (b2.south)	[]	{};
   \node [be]	at (b3.south)	[]	{};
   \node [be]	at (b4.south)	[]	{};
  \draw[thick,dotted]     ($(g2.north west)+(-0.1,0.08)$) rectangle ($(g2.south east)+(0.1,-0.8)$);
\end{tikzpicture}
}\subfloat{\hspace{1cm}
\begin{tikzpicture}[
    and/.style={and gate US,thick,draw,scale=0.8,fill=black!50,rotate=90,
		anchor=east},
    or/.style={or gate US,thick,draw,scale=0.8,fill=black!20,rotate=90,
		anchor=east},
    be/.style={circle,thick,draw,fill=yellow!90!red,anchor=north,
		minimum width=0.45cm},
    tr/.style={buffer gate US,thick,draw,fill=purple!60,rotate=90,
		anchor=east,minimum width=0.8cm},
    label distance=1mm,
    every label/.style={blue},
    event/.style={rectangle,thick,draw,fill=yellow!20,text width=0.7cm,
		text centered,font=\sffamily,anchor=north},
    edge from parent/.style={very thick,draw=black!70},
    edge from parent path={(\tikzparentnode.south) -- ++(0,-0.4cm)
			-| (\tikzchildnode.north)},
    level 1/.style={sibling distance=3cm,level distance=0.56cm,
			growth parent anchor=south,nodes=event},
    level 2/.style={sibling distance=2cm},
    level 3/.style={sibling distance=1.5cm},
    level 4/.style={sibling distance=1cm}
    ]
    \node (g1) [event] {\scriptsize $T$}
    child{
        child{ node(g2) {\scriptsize $G_1$}
        child{ 
        	child{ node(b1) {\scriptsize $B_1$} }
            child{ node(b2) {\scriptsize $B_2$} }
        }
        }
        child{ node(g3) {\scriptsize $G_2$}
        child{ 
          child{ node(b3) {\scriptsize $B_3$} }
          child{ node(b4) {\scriptsize $B_4$} }
        }
        }
    }
        ;
   \node [and]	at (g1.south)	[]	{};
   \node [or]	at (g2.south)	[]	{};
   \node [or]	at (g3.south)	[]	{};
   \node [be]	at (b1.south)	[]	{};
   \node [be]	at (b2.south)	[]	{};
   \node [be]	at (b3.south)	[]	{};
   \node [be]	at (b4.south)	[]	{};
  \draw[thick,dotted]     ($(g2.north west)+(-0.1,0.08)$) rectangle ($(g2.south east)+(0.1,-0.8)$);
\end{tikzpicture}
}
\centering
\caption{Example of \textit{G-mutate}: mutating gate $G_1$ into an OR gate.}
\label{fig:g-mutate}
\end{figure}
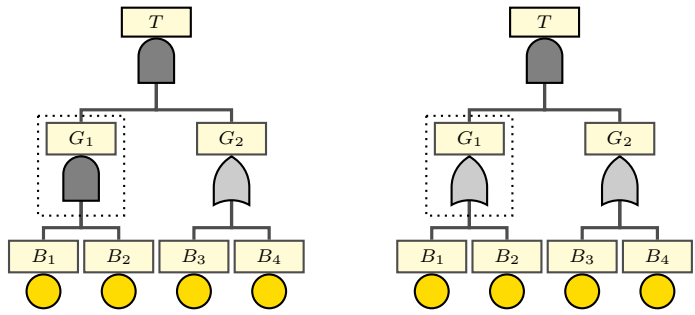

\begin{figure}[!ht]
\subfloat{
\begin{tikzpicture}[
    and/.style={and gate US,thick,draw,scale=0.8,fill=black!50,rotate=90,
		anchor=east},
    or/.style={or gate US,thick,draw,scale=0.8,fill=black!20,rotate=90,
		anchor=east},
    be/.style={circle,thick,draw,fill=yellow!90!red,anchor=north,
		minimum width=0.45cm},
    tr/.style={buffer gate US,thick,draw,fill=purple!60,rotate=90,
		anchor=east,minimum width=0.8cm},
    label distance=1mm,
    every label/.style={blue},
    event/.style={rectangle,thick,draw,fill=yellow!20,text width=0.7cm,
		text centered,font=\sffamily,anchor=north},
    edge from parent/.style={very thick,draw=black!70},
    edge from parent path={(\tikzparentnode.south) -- ++(0,-0.4cm)
			-| (\tikzchildnode.north)},
    level 1/.style={sibling distance=3cm,level distance=0.56cm,
			growth parent anchor=south,nodes=event},
    level 2/.style={sibling distance=1.5cm},
    level 3/.style={sibling distance=1.5cm},
    level 4/.style={sibling distance=1cm}
    ]
    \node (g1) [event] {\scriptsize $T$}
    child{
        child{ node(g2) {\scriptsize $G$}
        child{ 
        	child{ node(b1) {\scriptsize $B_1$} }
            child{ node(b2) {\scriptsize $B_2$} }
        }
        }
        child{ node(b3) {\scriptsize $B_3$} }
        child{ node(b4) {\scriptsize $B_4$} }
    }
        ;
   \node [and]	at (g1.south)	[]	{};
   \node [or]	at (g2.south)	[]	{};
   \node [be]	at (b1.south)	[]	{};
   \node [be]	at (b2.south)	[]	{};
   \node [be]	at (b3.south)	[]	{};
   \node [be]	at (b4.south)	[]	{};
   \draw[thick,dotted]     ($(g2.north west)+(-0.2,0.2)$) rectangle ($(g2.south east)+(0.2,-0.90)$);
\end{tikzpicture}
}\subfloat{\hspace{1cm}
\begin{tikzpicture}[
    and/.style={and gate US,thick,draw,scale=0.8,fill=black!50,rotate=90,
		anchor=east},
    or/.style={or gate US,thick,draw,scale=0.8,fill=black!20,rotate=90,
		anchor=east},
    be/.style={circle,thick,draw,fill=yellow!90!red,anchor=north,
		minimum width=0.45cm},
    tr/.style={buffer gate US,thick,draw,fill=purple!60,rotate=90,
		anchor=east,minimum width=0.8cm},
    label distance=1mm,
    every label/.style={blue},
    event/.style={rectangle,thick,draw,fill=yellow!20,text width=0.7cm,
		text centered,font=\sffamily,anchor=north},
    edge from parent/.style={very thick,draw=black!70},
    edge from parent path={(\tikzparentnode.south) -- ++(0,-0.4cm)
			-| (\tikzchildnode.north)},
    level 1/.style={sibling distance=1cm,level distance=0.56cm,
			growth parent anchor=south,nodes=event},
    level 2/.style={sibling distance=1cm},
    level 3/.style={sibling distance=1cm},
    level 4/.style={sibling distance=1cm}
    ]
    \node (g1) [event] {\scriptsize $T$}
    child{
        	child{ node(b1) {\scriptsize $B_1$} }
            child{ node(b2) {\scriptsize $B_2$} }
        child{ node(b3) {\scriptsize $B_3$} }
        child{ node(b4) {\scriptsize $B_4$} }
    }
        ;
   \node [and]	at (g1.south)	[]	{};
   \node [be]	at (b1.south)	[]	{};
   \node [be]	at (b2.south)	[]	{};
   \node [be]	at (b3.south)	[]	{};
   \node [be]	at (b4.south)	[]	{};
\end{tikzpicture}
}
\centering
\caption{Example of \textit{G-delete}: deleting gate $G$.}
\label{fig:g-delete}
\end{figure}
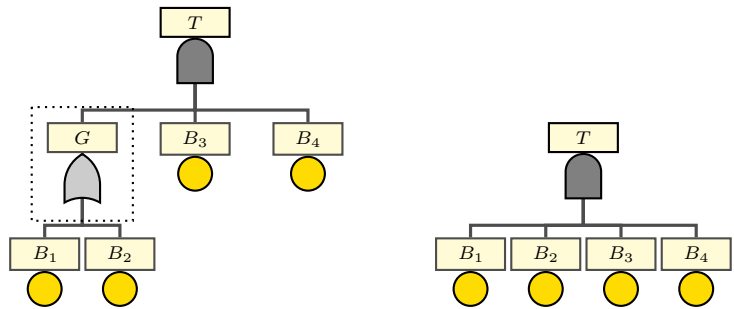

 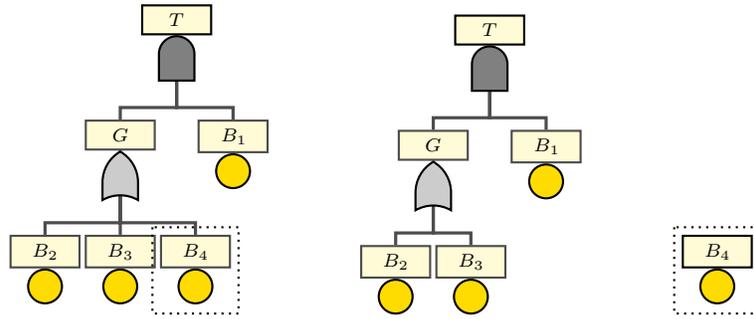
\begin{figure}[!ht]
\subfloat{
 \begin{tikzpicture}[
     and/.style={and gate US,thick,draw,scale=0.8,fill=black!50,rotate=90,
 		anchor=east},
     or/.style={or gate US,thick,draw,scale=0.8,fill=black!20,rotate=90,
 		anchor=east},
     be/.style={circle,thick,draw,fill=yellow!90!red,anchor=north,
 		minimum width=0.45cm},
     tr/.style={buffer gate US,thick,draw,fill=purple!60,rotate=90,
 		anchor=east,minimum width=0.8cm},
     label distance=1mm,
     every label/.style={blue},
     event/.style={rectangle,thick,draw,fill=yellow!20,text width=0.7cm,
 		text centered,font=\sffamily,anchor=north},
     edge from parent/.style={very thick,draw=black!70},
     edge from parent path={(\tikzparentnode.south) -- ++(0,-0.4cm)
 			-| (\tikzchildnode.north)},
     level 1/.style={sibling distance=3cm,level distance=0.56cm,
 			growth parent anchor=south,nodes=event},
     level 2/.style={sibling distance=1.5cm},
     level 3/.style={sibling distance=1.5cm},
     level 4/.style={sibling distance=1cm}
     ]
     \node (g1) [event] {\scriptsize $T$}
     child{
         child{ node(g2) {\scriptsize $G$}
         child{ 
         	child{ node(b2) {\scriptsize $B_2$} }
             child{ node(b3) {\scriptsize $B_3$} }
             child{ node(b4) {\scriptsize $B_4$} }
         }
         }
         child{ node(b1) {\scriptsize $B_1$} }
     }
         ;
\node [and]	at (g1.south)	[]	{};
    \node [or]	at (g2.south)	[]	{};
    \node [be]	at (b1.south)	[]	{};
    \node [be]	at (b2.south)	[]	{};
    \node [be]	at (b3.south)	[]	{};
    \node [be]	at (b4.south)	[]	{};
    \draw[thick,dotted]     ($(b4.north west)+(-0.1,0.1)$) rectangle ($(b4.south east)+(0.1,-0.6)$);
 \end{tikzpicture}
 }\subfloat{\hspace{1cm}
 \begin{tikzpicture}[
     and/.style={and gate US,thick,draw,scale=0.8,fill=black!50,rotate=90,
 		anchor=east},
     or/.style={or gate US,thick,draw,scale=0.8,fill=black!20,rotate=90,
 		anchor=east},
     be/.style={circle,thick,draw,fill=yellow!90!red,anchor=north,
 		minimum width=0.45cm},
     tr/.style={buffer gate US,thick,draw,fill=purple!60,rotate=90,
 		anchor=east,minimum width=0.8cm},
     label distance=1mm,
     every label/.style={blue},
     event/.style={rectangle,thick,draw,fill=yellow!20,text width=0.7cm,
 		text centered,font=\sffamily,anchor=north},
     edge from parent/.style={very thick,draw=black!70},
     edge from parent path={(\tikzparentnode.south) -- ++(0,-0.4cm)
 			-| (\tikzchildnode.north)},
     level 1/.style={sibling distance=3cm,level distance=0.56cm,
 			growth parent anchor=south,nodes=event},
     level 2/.style={sibling distance=1.5cm},
     level 3/.style={sibling distance=1.5cm},
     level 4/.style={sibling distance=1cm}
     ]
     \node (g1) [event] {\scriptsize $T$}
     child{
         child{ node(g2) {\scriptsize $G$}
         child{ 
         	child{ node(b2) {\scriptsize $B_2$} }
             child{ node(b3) {\scriptsize $B_3$} }
         }
         }
         child{ node(b1) {\scriptsize $B_1$} }
     }
         ;
    \node [and]	at (g1.south)	[]	{};
    \node [or]	at (g2.south)	[]	{};
    \node [be]	at (b1.south)	[]	{};
    \node [be]	at (b2.south)	[]	{};
    \node [be]	at (b3.south)	[]	{};
 \end{tikzpicture}
 }\subfloat{\hspace{1cm}
 \begin{tikzpicture}[
     and/.style={and gate US,thick,draw,scale=0.8,fill=black!50,rotate=90,
 		anchor=east},
     or/.style={or gate US,thick,draw,scale=0.8,fill=black!20,rotate=90,
 		anchor=east},
     be/.style={circle,thick,draw,fill=yellow!90!red,anchor=north,
 		minimum width=0.45cm},
     tr/.style={buffer gate US,thick,draw,fill=purple!60,rotate=90,
 		anchor=east,minimum width=0.8cm},
     label distance=1mm,
     every label/.style={blue},
     event/.style={rectangle,thick,draw,fill=yellow!20,text width=0.7cm,
 		text centered,font=\sffamily,anchor=north},
     edge from parent/.style={very thick,draw=black!70},
     edge from parent path={(\tikzparentnode.south) -- ++(0,-0.4cm)
 			-| (\tikzchildnode.north)},
     level 1/.style={sibling distance=3cm,level distance=0.56cm,
 			growth parent anchor=south,nodes=event},
     level 2/.style={sibling distance=1.5cm},
     level 3/.style={sibling distance=1.5cm},
     level 4/.style={sibling distance=1cm}
     ]
     \node (g1) [event] {\scriptsize $B_4$}
         ;
    \node [be]	at (g1.south)	[]	{};
    \draw[thick,dotted]     ($(g1.north west)+(-0.1,0.1)$) rectangle ($(g1.south east)+(0.1,-0.6)$);
 \end{tikzpicture}
 }
 \centering
 \caption{Disconnecting basic event $B_4$ from a FT.}
 \label{fig:BE-disconnect}
 \end{figure}

\newpage

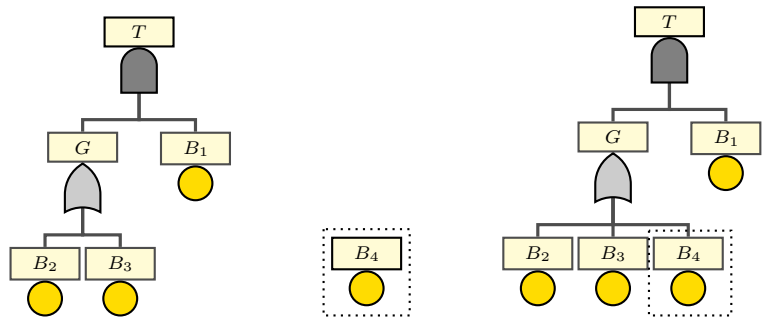
\begin{figure}[!ht]
\subfloat{
\begin{tikzpicture}[
    and/.style={and gate US,thick,draw,scale=0.8,fill=black!50,rotate=90,
		anchor=east},
    or/.style={or gate US,thick,draw,scale=0.8,fill=black!20,rotate=90,
		anchor=east},
    be/.style={circle,thick,draw,fill=yellow!90!red,anchor=north,
		minimum width=0.45cm},
    tr/.style={buffer gate US,thick,draw,fill=purple!60,rotate=90,
		anchor=east,minimum width=0.8cm},
    label distance=1mm,
    every label/.style={blue},
    event/.style={rectangle,thick,draw,fill=yellow!20,text width=0.7cm,
		text centered,font=\sffamily,anchor=north},
    edge from parent/.style={very thick,draw=black!70},
    edge from parent path={(\tikzparentnode.south) -- ++(0,-0.4cm)
			-| (\tikzchildnode.north)},
    level 1/.style={sibling distance=3cm,level distance=0.56cm,
			growth parent anchor=south,nodes=event},
    level 2/.style={sibling distance=1.5cm},
    level 3/.style={sibling distance=1.5cm},
    level 4/.style={sibling distance=1cm}
    ]
    \node (g1) [event] {\scriptsize $T$}
    child{
        child{ node(g2) {\scriptsize $G$}
        child{ 
        	child{ node(b2) {\scriptsize $B_2$} }
            child{ node(b3) {\scriptsize $B_3$} }
        }
        }
        child{ node(b1) {\scriptsize $B_1$} }
    }
        ;
   \node [and]	at (g1.south)	[]	{};
   \node [or]	at (g2.south)	[]	{};
   \node [be]	at (b1.south)	[]	{};
   \node [be]	at (b2.south)	[]	{};
   \node [be]	at (b3.south)	[]	{};
\end{tikzpicture}
}\subfloat{\hspace{1cm}
\begin{tikzpicture}[
    and/.style={and gate US,thick,draw,scale=0.8,fill=black!50,rotate=90,
		anchor=east},
    or/.style={or gate US,thick,draw,scale=0.8,fill=black!20,rotate=90,
		anchor=east},
    be/.style={circle,thick,draw,fill=yellow!90!red,anchor=north,
		minimum width=0.45cm},
    tr/.style={buffer gate US,thick,draw,fill=purple!60,rotate=90,
		anchor=east,minimum width=0.8cm},
    label distance=1mm,
    every label/.style={blue},
    event/.style={rectangle,thick,draw,fill=yellow!20,text width=0.7cm,
		text centered,font=\sffamily,anchor=north},
    edge from parent/.style={very thick,draw=black!70},
    edge from parent path={(\tikzparentnode.south) -- ++(0,-0.4cm)
			-| (\tikzchildnode.north)},
    level 1/.style={sibling distance=3cm,level distance=0.56cm,
			growth parent anchor=south,nodes=event},
    level 2/.style={sibling distance=1.5cm},
    level 3/.style={sibling distance=1.5cm},
    level 4/.style={sibling distance=1cm}
    ]
    \node (g1) [event] {\scriptsize $B_4$}
        ;
   \node [be]	at (g1.south)	[]	{};
   \draw[thick,dotted]     ($(g1.north west)+(-0.1,0.1)$) rectangle ($(g1.south east)+(0.1,-0.6)$);
\end{tikzpicture}
}\subfloat{\hspace{1cm}
\begin{tikzpicture}[
    and/.style={and gate US,thick,draw,scale=0.8,fill=black!50,rotate=90,
		anchor=east},
    or/.style={or gate US,thick,draw,scale=0.8,fill=black!20,rotate=90,
		anchor=east},
    be/.style={circle,thick,draw,fill=yellow!90!red,anchor=north,
		minimum width=0.45cm},
    tr/.style={buffer gate US,thick,draw,fill=purple!60,rotate=90,
		anchor=east,minimum width=0.8cm},
    label distance=1mm,
    every label/.style={blue},
    event/.style={rectangle,thick,draw,fill=yellow!20,text width=0.7cm,
		text centered,font=\sffamily,anchor=north},
    edge from parent/.style={very thick,draw=black!70},
    edge from parent path={(\tikzparentnode.south) -- ++(0,-0.4cm)
			-| (\tikzchildnode.north)},
    level 1/.style={sibling distance=3cm,level distance=0.56cm,
			growth parent anchor=south,nodes=event},
    level 2/.style={sibling distance=1.5cm},
    level 3/.style={sibling distance=1.5cm},
    level 4/.style={sibling distance=1cm}
    ]
    \node (g1) [event] {\scriptsize $T$}
    child{
        child{ node(g2) {\scriptsize $G$}
        child{ 
        	child{ node(b2) {\scriptsize $B_2$} }
            child{ node(b3) {\scriptsize $B_3$} }
            child{ node(b4) {\scriptsize $B_4$} }
        }
        }
        child{ node(b1) {\scriptsize $B_1$} }
    }
        ;
   \node [and]	at (g1.south)	[]	{};
   \node [or]	at (g2.south)	[]	{};
   \node [be]	at (b1.south)	[]	{};
   \node [be]	at (b2.south)	[]	{};
   \node [be]	at (b3.south)	[]	{};
   \node [be]	at (b4.south)	[]	{};
   \draw[thick,dotted]     ($(b4.north west)+(-0.05,0.08)$) rectangle ($(b4.south east)+(0.1,-0.6)$);
\end{tikzpicture}
}
\centering
\caption{Example of \textit{BE-connect}: connecting $B_4$ to a FT.}
\label{fig:BE-connect}
\end{figure}

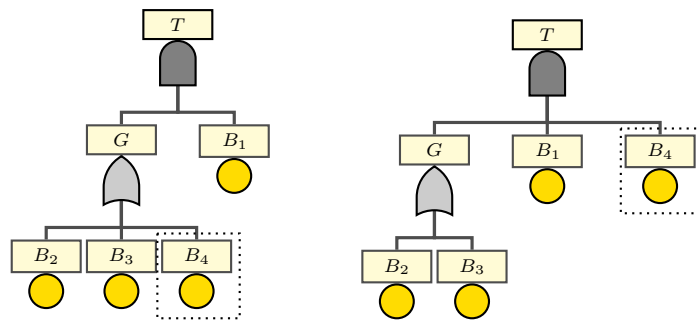
\begin{figure}[!ht]
\subfloat{
\begin{tikzpicture}[
    and/.style={and gate US,thick,draw,scale=0.8,fill=black!50,rotate=90,
		anchor=east},
    or/.style={or gate US,thick,draw,scale=0.8,fill=black!20,rotate=90,
		anchor=east},
    be/.style={circle,thick,draw,fill=yellow!90!red,anchor=north,
		minimum width=0.45cm},
    tr/.style={buffer gate US,thick,draw,fill=purple!60,rotate=90,
		anchor=east,minimum width=0.8cm},
    label distance=1mm,
    every label/.style={blue},
    event/.style={rectangle,thick,draw,fill=yellow!20,text width=0.7cm,
		text centered,font=\sffamily,anchor=north},
    edge from parent/.style={very thick,draw=black!70},
    edge from parent path={(\tikzparentnode.south) -- ++(0,-0.4cm)
			-| (\tikzchildnode.north)},
    level 1/.style={sibling distance=3cm,level distance=0.56cm,
			growth parent anchor=south,nodes=event},
    level 2/.style={sibling distance=1.5cm},
    level 3/.style={sibling distance=1.5cm},
    level 4/.style={sibling distance=1cm}
    ]
    \node (g1) [event] {\scriptsize $T$}
    child{
        child{ node(g2) {\scriptsize $G$}
        child{ 
        	child{ node(b2) {\scriptsize $B_2$} }
            child{ node(b3) {\scriptsize $B_3$} }
            child{ node(b4) {\scriptsize $B_4$} }
        }
        }
        child{ node(b1) {\scriptsize $B_1$} }
    }
        ;
   \node [and]	at (g1.south)	[]	{};
   \node [or]	at (g2.south)	[]	{};
   \node [be]	at (b1.south)	[]	{};
   \node [be]	at (b2.south)	[]	{};
   \node [be]	at (b3.south)	[]	{};
   \node [be]	at (b4.south)	[]	{};
   \draw[thick,dotted]     ($(b4.north west)+(-0.05,0.08)$) rectangle ($(b4.south east)+(0.1,-0.6)$);
\end{tikzpicture}
}\subfloat{\hspace{1cm}
\begin{tikzpicture}[
    and/.style={and gate US,thick,draw,scale=0.8,fill=black!50,rotate=90,
		anchor=east},
    or/.style={or gate US,thick,draw,scale=0.8,fill=black!20,rotate=90,
		anchor=east},
    be/.style={circle,thick,draw,fill=yellow!90!red,anchor=north,
		minimum width=0.45cm},
    tr/.style={buffer gate US,thick,draw,fill=purple!60,rotate=90,
		anchor=east,minimum width=0.8cm},
    label distance=1mm,
    every label/.style={blue},
    event/.style={rectangle,thick,draw,fill=yellow!20,text width=0.7cm,
		text centered,font=\sffamily,anchor=north},
    edge from parent/.style={very thick,draw=black!70},
    edge from parent path={(\tikzparentnode.south) -- ++(0,-0.4cm)
			-| (\tikzchildnode.north)},
    level 1/.style={sibling distance=3cm,level distance=0.56cm,
			growth parent anchor=south,nodes=event},
    level 2/.style={sibling distance=1.5cm},
    level 3/.style={sibling distance=1.5cm},
    level 4/.style={sibling distance=1cm}
    ]
    \node (g1) [event] {\scriptsize $T$}
    child{
        child{ node(g2) {\scriptsize $G$}
        child{ 
        	child{ node(b2) {\scriptsize $B_2$} }
            child{ node(b3) {\scriptsize $B_3$} }
        }
        }
        child{ node(b1) {\scriptsize $B_1$} }
        child{ node(b4) {\scriptsize $B_4$} }
    }
        ;
   \node [and]	at (g1.south)	[]	{};
   \node [or]	at (g2.south)	[]	{};
   \node [be]	at (b1.south)	[]	{};
   \node [be]	at (b2.south)	[]	{};
   \node [be]	at (b3.south)	[]	{};
   \node [be]	at (b4.south)	[]	{};
  \draw[thick,dotted]     ($(b4.north west)+(-0.05,0.08)$) rectangle ($(b4.south east)+(0.1,-0.6)$);
\end{tikzpicture}
}
\centering
\caption{Example of \textit{BE-swap}: swapping basic event $B_4$ from gate $G$ to gate $T$.}
\label{fig:BE-swap}
\end{figure}

\newpage

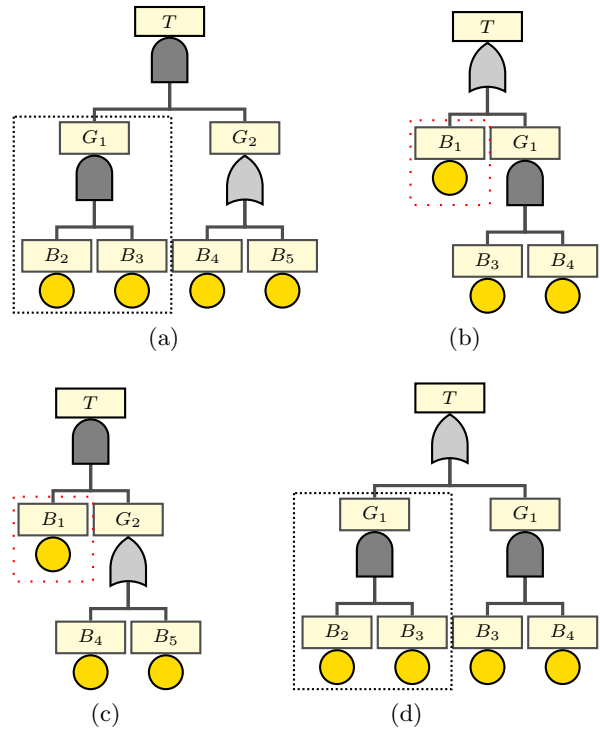
\begin{figure}[!ht]
\vspace*{3cm}
\centering
\subfloat[]{
\begin{tikzpicture}[
    and/.style={and gate US,thick,draw,scale=0.8,fill=black!50,rotate=90,
		anchor=east},
    or/.style={or gate US,thick,draw,scale=0.8,fill=black!20,rotate=90,
		anchor=east},
    be/.style={circle,thick,draw,fill=yellow!90!red,anchor=north,
		minimum width=0.45cm},
    tr/.style={buffer gate US,thick,draw,fill=purple!60,rotate=90,
		anchor=east,minimum width=0.8cm},
    label distance=1mm,
    every label/.style={blue},
    event/.style={rectangle,thick,draw,fill=yellow!20,text width=0.7cm,
		text centered,font=\sffamily,anchor=north},
    edge from parent/.style={very thick,draw=black!70},
    edge from parent path={(\tikzparentnode.south) -- ++(0,-0.4cm)
			-| (\tikzchildnode.north)},
    level 1/.style={sibling distance=3cm,level distance=0.56cm,
			growth parent anchor=south,nodes=event},
    level 2/.style={sibling distance=2cm},
    level 3/.style={sibling distance=1.5cm},
    level 4/.style={sibling distance=1cm}
    ]
    \node (g1) [event] {\scriptsize $T$}
    child{
        child{ node(g2) {\scriptsize $G_1$}
        child{ 
        	child{ node(b1) {\scriptsize $B_2$} }
            child{ node(b2) {\scriptsize $B_3$} }
        }
        }
        child{ node(g3) {\scriptsize $G_2$}
        child{ 
          child{ node(b3) {\scriptsize $B_4$} }
          child{ node(b4) {\scriptsize $B_5$} }
        }
        }
    }
        ;
   \node [and]	at (g1.south)	[]	{};
   \node [and]	at (g2.south)	[]	{};
   \node [or]	at (g3.south)	[]	{};
   \node [be]	at (b1.south)	[]	{};
   \node [be]	at (b2.south)	[]	{};
   \node [be]	at (b3.south)	[]	{};
   \node [be]	at (b4.south)	[]	{};
   \draw[thick,densely dotted]     ($(g2.north west)+(-0.6,0.06)$) rectangle ($(g2.south east)+(0.55,-2.1)$);
\end{tikzpicture}
}\subfloat[]{\hspace{1cm}
\begin{tikzpicture}[
    and/.style={and gate US,thick,draw,scale=0.8,fill=black!50,rotate=90,
		anchor=east},
    or/.style={or gate US,thick,draw,scale=0.8,fill=black!20,rotate=90,
		anchor=east},
    be/.style={circle,thick,draw,fill=yellow!90!red,anchor=north,
		minimum width=0.45cm},
    tr/.style={buffer gate US,thick,draw,fill=purple!60,rotate=90,
		anchor=east,minimum width=0.8cm},
    label distance=1mm,
    every label/.style={blue},
    event/.style={rectangle,thick,draw,fill=yellow!20,text width=0.7cm,
		text centered,font=\sffamily,anchor=north},
    edge from parent/.style={very thick,draw=black!70},
    edge from parent path={(\tikzparentnode.south) -- ++(0,-0.4cm)
			-| (\tikzchildnode.north)},
    level 1/.style={sibling distance=3cm,level distance=0.56cm,
			growth parent anchor=south,nodes=event},
    level 2/.style={sibling distance=1cm},
    level 3/.style={sibling distance=1.5cm},
    level 4/.style={sibling distance=1cm}
    ]
    \node (g1) [event] {\scriptsize $T$}
    child{
    	child{ node(b1) {\scriptsize $B_1$}  }
        child{ node(g2) {\scriptsize $G_1$}
        child{ 
        	child{ node(b2) {\scriptsize $B_3$} }
            child{ node(b3) {\scriptsize $B_4$} }
        }
        }
    }
        ;
   \node [or]	at (g1.south)	[]	{};
   \node [and]	at (g2.south)	[]	{};
   \node [be]	at (b1.south)	[]	{};
   \node [be]	at (b2.south)	[]	{};
   \node [be]	at (b3.south)	[]	{};
   \draw[red,thick,loosely dotted]     ($(b1.north west)+(-0.05,0.08)$) rectangle ($(b1.south east)+(0.06,-0.6)$);
\end{tikzpicture}
}

\subfloat[]{
\begin{tikzpicture}[
    and/.style={and gate US,thick,draw,scale=0.8,fill=black!50,rotate=90,
		anchor=east},
    or/.style={or gate US,thick,draw,scale=0.8,fill=black!20,rotate=90,
		anchor=east},
    be/.style={circle,thick,draw,fill=yellow!90!red,anchor=north,
		minimum width=0.45cm},
    tr/.style={buffer gate US,thick,draw,fill=purple!60,rotate=90,
		anchor=east,minimum width=0.8cm},
    label distance=1mm,
    every label/.style={blue},
    event/.style={rectangle,thick,draw,fill=yellow!20,text width=0.7cm,
		text centered,font=\sffamily,anchor=north},
    edge from parent/.style={very thick,draw=black!70},
    edge from parent path={(\tikzparentnode.south) -- ++(0,-0.4cm)
			-| (\tikzchildnode.north)},
    level 1/.style={sibling distance=3cm,level distance=0.56cm,
			growth parent anchor=south,nodes=event},
    level 2/.style={sibling distance=1cm},
    level 3/.style={sibling distance=1.5cm},
    level 4/.style={sibling distance=1cm}
    ]
    \node (g1) [event] {\scriptsize $T$}
    child{
        child{ node(b1) {\scriptsize $B_1$}  }
        child{ node(g3) {\scriptsize $G_2$}
        child{ 
          child{ node(b3) {\scriptsize $B_4$} }
          child{ node(b4) {\scriptsize \scriptsize $B_5$} }
        }
        }
    }
        ;
   \node [and]	at (g1.south)	[]	{};
   \node [or]	at (g3.south)	[]	{};
   \node [be]	at (b1.south)	[]	{};
   \node [be]	at (b3.south)	[]	{};
   \node [be]	at (b4.south)	[]	{};
   \draw[red,thick,loosely dotted]     ($(b1.north west)+(-0.05,0.08)$) rectangle ($(b1.south east)+(0.06,-0.6)$);
\end{tikzpicture}
}\subfloat[]{\hspace{1cm}
\begin{tikzpicture}[
    and/.style={and gate US,thick,draw,scale=0.8,fill=black!50,rotate=90,
		anchor=east},
    or/.style={or gate US,thick,draw,scale=0.8,fill=black!20,rotate=90,
		anchor=east},
    be/.style={circle,thick,draw,fill=yellow!90!red,anchor=north,
		minimum width=0.45cm},
    tr/.style={buffer gate US,thick,draw,fill=purple!60,rotate=90,
		anchor=east,minimum width=0.8cm},
    label distance=1mm,
    every label/.style={blue},
    event/.style={rectangle,thick,draw,fill=yellow!20,text width=0.7cm,
		text centered,font=\sffamily,anchor=north},
    edge from parent/.style={very thick,draw=black!70},
    edge from parent path={(\tikzparentnode.south) -- ++(0,-0.4cm)
			-| (\tikzchildnode.north)},
    level 1/.style={sibling distance=3cm,level distance=0.56cm,
			growth parent anchor=south,nodes=event},
    level 2/.style={sibling distance=2cm},
    level 3/.style={sibling distance=1.5cm},
    level 4/.style={sibling distance=1cm}
    ]
    \node (g1) [event] {\scriptsize $T$}
    child{
    	child{ node(g2) {\scriptsize $G_1$}
        child{ 
        	child{ node(b1) {\scriptsize $B_2$} }
            child{ node(b2) {\scriptsize $B_3$} }
        }
        }
        child{ node(g3) {\scriptsize $G_1$}
        child{ 
        	child{ node(b3) {\scriptsize $B_3$} }
            child{ node(b4) {\scriptsize $B_4$} }
        }
        }
    }
        ;
   \node [or]	at (g1.south)	[]	{};
   \node [and]	at (g2.south)	[]	{};
   \node [and]	at (g3.south)	[]	{};
   \node [be]	at (b1.south)	[]	{};
   \node [be]	at (b2.south)	[]	{};
   \node [be]	at (b3.south)	[]	{};
   \node [be]	at (b4.south)	[]	{};
   \draw[thick,densely dotted]     ($(g2.north west)+(-0.6,0.06)$) rectangle ($(g2.south east)+(0.55,-2.1)$);
\end{tikzpicture}
}
\centering
\caption{Example of \textit{crossover} between (a) and (b) and resulting children (c and d).}
\label{fig:crossover}
\end{figure}

\newpage

\section{Selection Strategies}
\label{app:selection-strategies}

In this section, we compare different selections strategies for our evolutionary algorithms:
\begin{itemize}
    \item \textit{Elitism}: systematically selects the best fitted individuals. This guarantees that the solution quality obtained by the EA will not decrease from one generation to the next.
    \item \textit{Roulette wheel Selection}: also known as \textit{fitness proportionate selection}, individuals in the population will be selected proportional to their fitnesses.
    \item \textit{Stochastic Universal Sampling} (SUS): chooses several individuals from the population by repeated random sampling, and uses a single random value to sample all of the individuals by choosing them at evenly spaced intervals. This allows least fitted individuals to be selected.
    \item \textit{Tournament Selection}: involves running several ``tournaments'' among a few individuals chosen at random from the population. The winner of each tournament (i.e. the individual with the best fitness) is selected.
    \item \textit{Random Selection}: the FTs are randomly selected in the population.
\end{itemize}

\begin{figure}[!ht]
\hspace*{-0.4cm}
\subfloat[Average FT accuracy during the EA.\label{fig:results-selection-strategies}]{
    \includegraphics[width=0.51\linewidth]{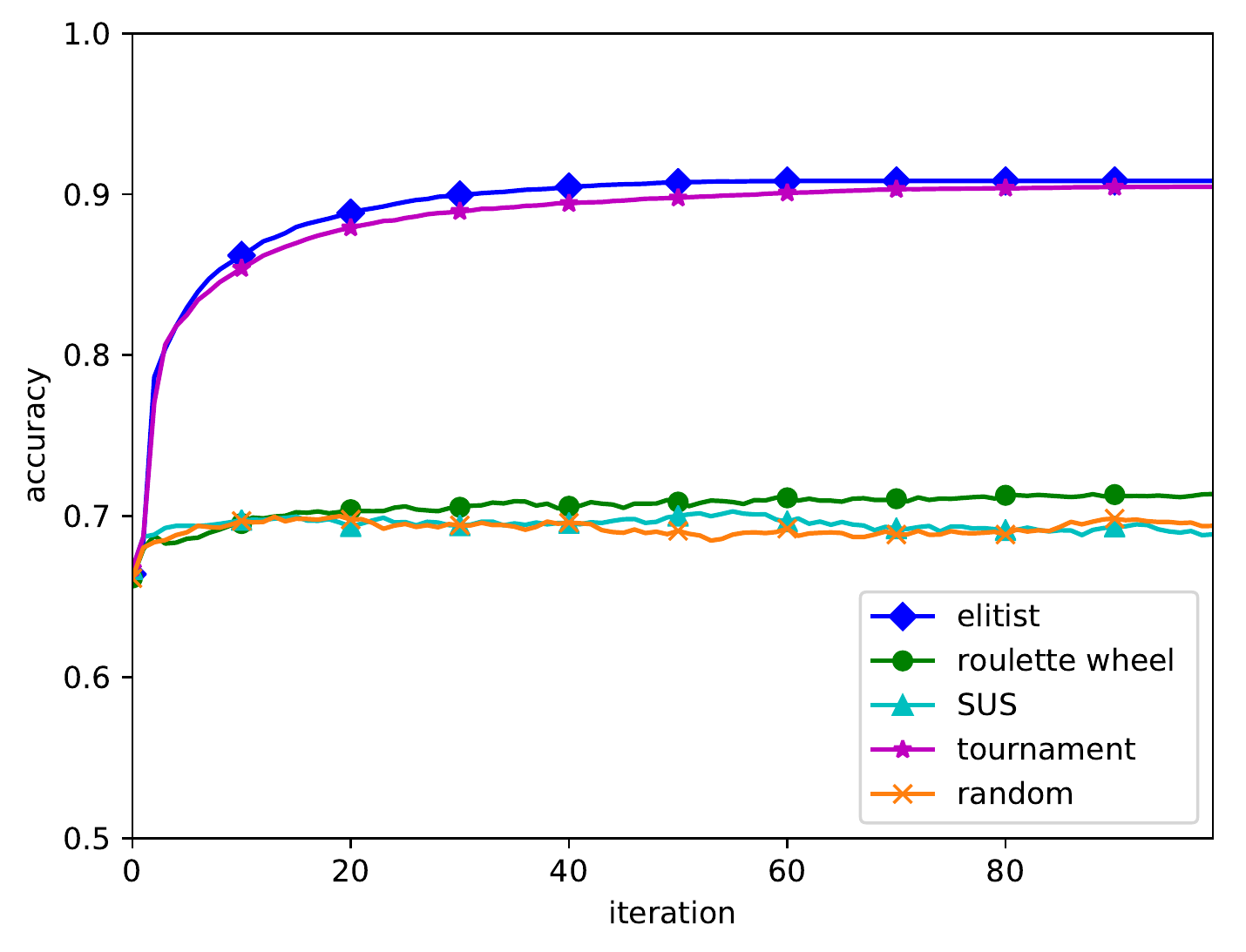}
}\subfloat[Average FT accuracy.\label{fig:results-selection-bes}]{
    \includegraphics[width=0.51\linewidth]{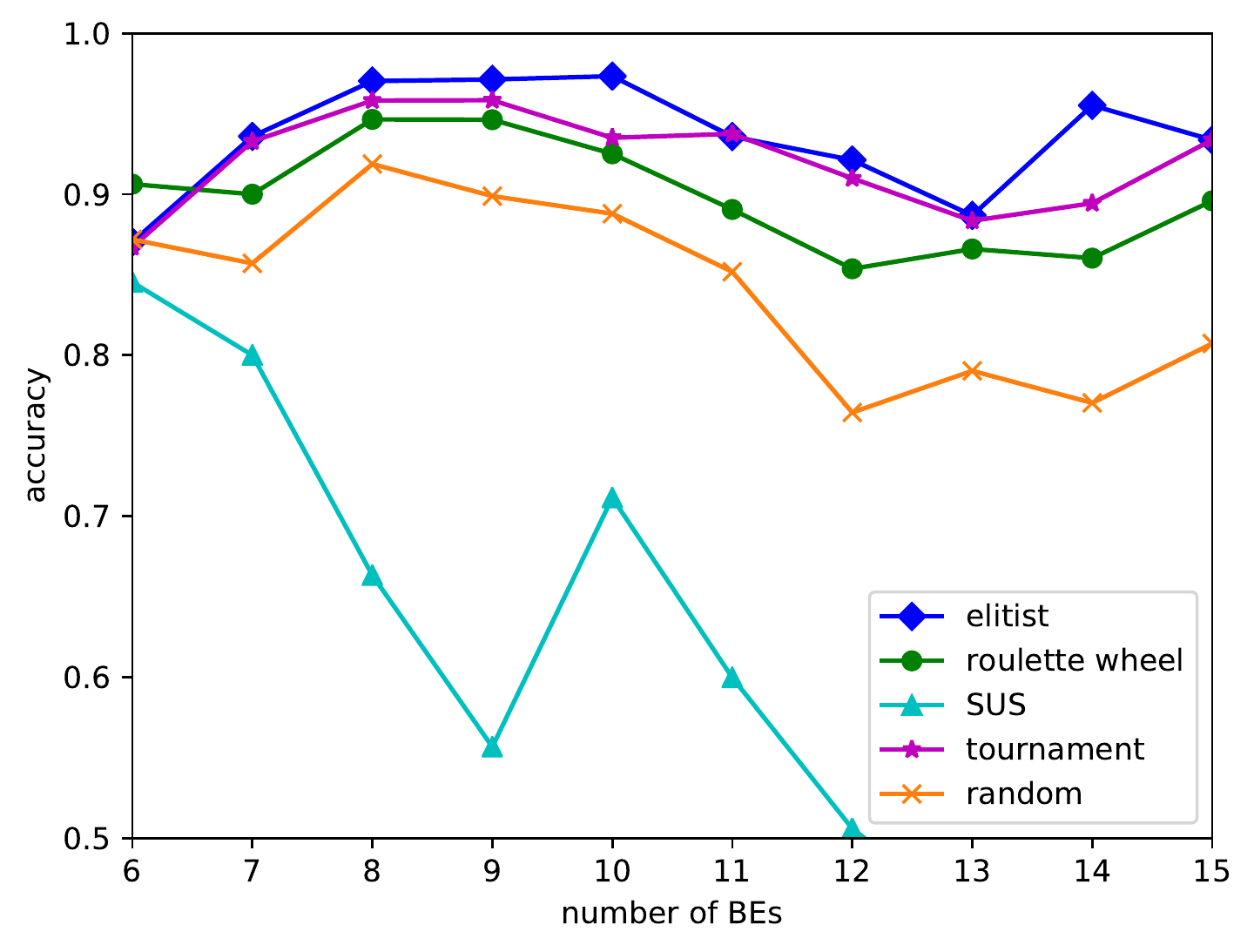}
}
\caption{Comparison of different selection strategies.}
\label{fig:comparison-strategies}
\end{figure}

The results of this experiment are shown in Fig. \ref{fig:comparison-strategies}. We show in Fig. \ref{fig:results-selection-strategies} how the accuracy of the best FT in the population evolves through iterations. In Fig. \ref{fig:results-selection-bes}, we show the average FT accuracy obtained as a function of the number of BEs in the FT.
These results show that the elitist strategy leads to the highest FT accuracy, and to the fastest convergence.

\newpage

\section{Industrial Case Study}
\label{app:industrial-case}

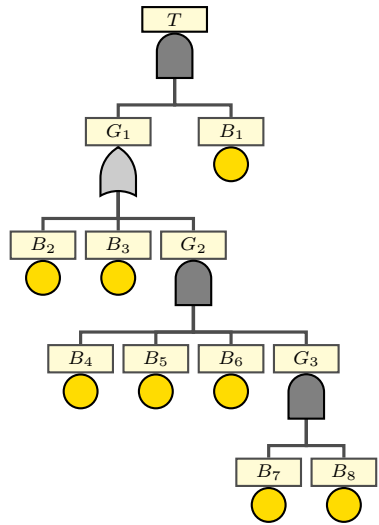
\begin{figure}[!ht]
\vspace*{3cm}
\begin{center}
\begin{tiny}
\begin{tikzpicture}[
    and/.style={and gate US,thick,draw,scale=0.8,fill=black!50,rotate=90,
		anchor=east},
    or/.style={or gate US,thick,draw,scale=0.8,fill=black!20,rotate=90,
		anchor=east},
    be/.style={circle,thick,draw,fill=yellow!90!red,anchor=north,
		minimum width=0.45cm},
    tr/.style={buffer gate US,thick,draw,fill=purple!60,rotate=90,
		anchor=east,minimum width=0.8cm},
    label distance=1mm,
    every label/.style={blue},
    event/.style={rectangle,thick,draw,fill=yellow!20,text width=0.7cm,
		text centered,font=\sffamily,anchor=north},
    edge from parent/.style={very thick,draw=black!70},
    edge from parent path={(\tikzparentnode.south) -- ++(0,-0.4cm)
			-| (\tikzchildnode.north)},
    level 1/.style={sibling distance=3cm,level distance=0.56cm,
			growth parent anchor=south,nodes=event},
    level 2/.style={sibling distance=1.5cm},
    level 3/.style={sibling distance=1.5cm},
    level 4/.style={sibling distance=1cm}
    ]
    \node (g1) [event] {\scriptsize $T$}
    child{
        child{ node(g2) {\scriptsize $G_1$}
        child{ 
        	child{ node(b1) {\scriptsize $B_2$} }
            child{ node(b2) {\scriptsize $B_3$} }
            child{ node(g3) {\scriptsize $G_2$}
              child{
                child{ node(b6) {\scriptsize $B_4$} }
                child{ node(b4) {\scriptsize $B_5$} }
                child{ node(b5) {\scriptsize $B_6$} }
                child{ node(g4) {\scriptsize $G_3$}
                  child{
                    child{ node(b7) {\scriptsize $B_7$} }
                    child{ node(b8) {\scriptsize $B_8$} }
                  }
                 }
              }
            }
        }
        }
        child{ node(b3) {\scriptsize $B_1$} }
        }
        ;
   \node [and]	at (g1.south)	[]	{};
   \node [or]	at (g2.south)	[]	{};
   \node [and]	at (g3.south)	[]	{};
   \node [be]	at (b1.south)	[]	{};
   \node [be]	at (b2.south)	[]	{};
   \node [be]	at (b3.south)	[]	{};
   \node [be]	at (b4.south)	[]	{};
   \node [be]	at (b5.south)	[]	{};
   \node [be]	at (b6.south)	[]	{};
   \node [and]	at (g4.south)	[]	{};
   \node [be]	at (b7.south)	[]	{};
   \node [be]	at (b8.south)	[]	{};
 
\end{tikzpicture}
\end{tiny}
\end{center}
\caption{Fault Tree for the nozzle case study learned by our evolutionary algorithm.}
\label{fig:ft-nozzle}
\end{figure}

\newpage

\section{DNF of Fault Trees}
\label{app:cnf-dnf}

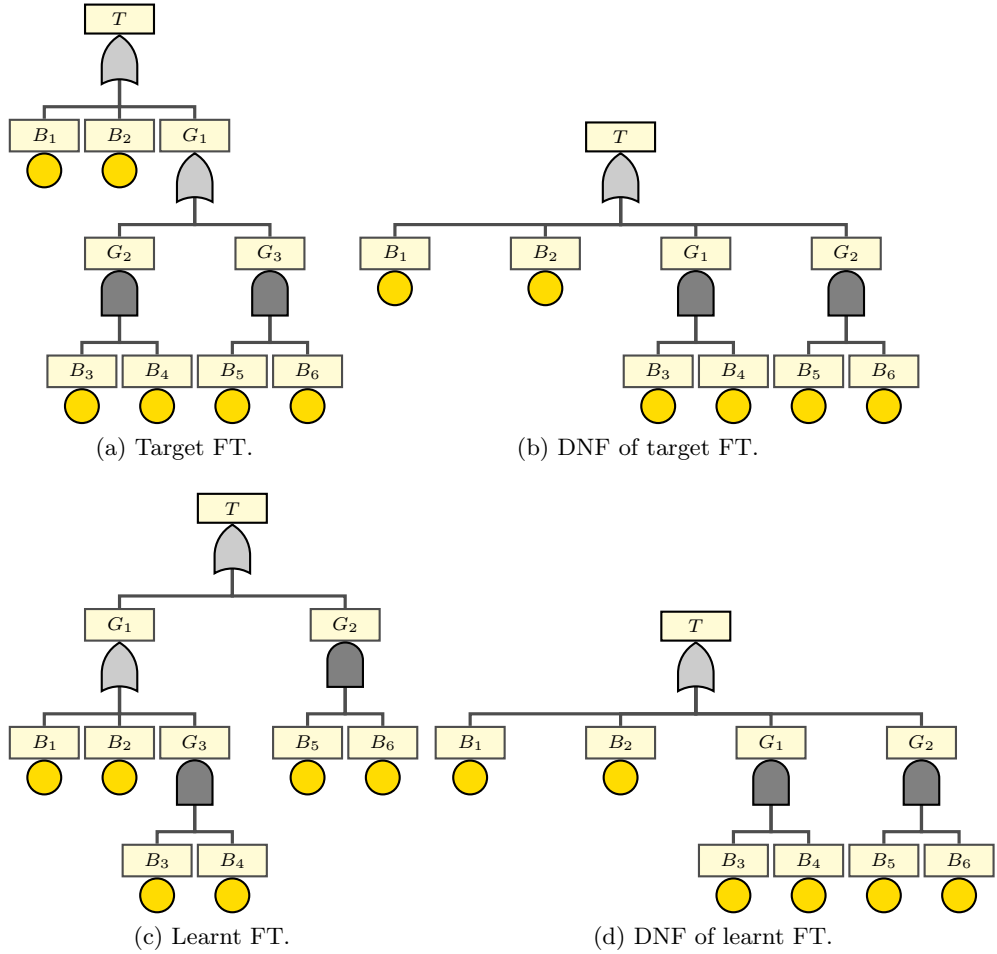
\begin{figure*}[ht]
\vspace*{1cm}
\subfloat[Target FT.]{
\begin{tikzpicture}[
    and/.style={and gate US,thick,draw,scale=0.8,fill=black!50,rotate=90,
		anchor=east},
    or/.style={or gate US,thick,draw,scale=0.8,fill=black!20,rotate=90,
		anchor=east},
    be/.style={circle,thick,draw,fill=yellow!90!red,anchor=north,
		minimum width=0.45cm},
    tr/.style={buffer gate US,thick,draw,fill=purple!60,rotate=90,
		anchor=east,minimum width=0.8cm},
    label distance=1mm,
    every label/.style={blue},
    event/.style={rectangle,thick,draw,fill=yellow!20,text width=0.7cm,
		text centered,font=\sffamily,anchor=north},
    edge from parent/.style={very thick,draw=black!70},
    edge from parent path={(\tikzparentnode.south) -- ++(0,-0.4cm)
			-| (\tikzchildnode.north)},
    level 1/.style={sibling distance=3cm,level distance=0.56cm,
			growth parent anchor=south,nodes=event},
    level 2/.style={sibling distance=1cm},
    level 3/.style={sibling distance=3cm},
    level 4/.style={sibling distance=2cm},
    level 5/.style={sibling distance=1cm}
    ]
    \node (g1) [event] {\scriptsize $T$}
    child{
        child{ node(b1) {\scriptsize $B_1$} }
        child{ node(b2) {\scriptsize $B_2$} }
        child{ node(g3) {\scriptsize $G_1$}
        child{ 
          child{ node(g2) {\scriptsize $G_2$}
            child{ 
              child{ node(b3) {\scriptsize $B_3$} }
              child{ node(b4) {\scriptsize $B_4$} }
            }
            }
            child{ node(g4) {\scriptsize $G_3$}
            child{ 
               child{ node(b5) {\scriptsize $B_5$} }
          	   child{ node(b6) {\scriptsize $B_6$} }
            }
            }
        }
        }
    }
        ;
   \node [or]	at (g1.south)	[]	{};
   \node [and]	at (g2.south)	[]	{};
   \node [and]	at (g4.south)	[]	{};
   \node [or]	at (g3.south)	[]	{};
   \node [be]	at (b1.south)	[]	{};
   \node [be]	at (b2.south)	[]	{};
   \node [be]	at (b3.south)	[]	{};
   \node [be]	at (b4.south)	[]	{};
   \node [be]	at (b5.south)	[]	{};
   \node [be]	at (b6.south)	[]	{};
\end{tikzpicture}
}
\subfloat[DNF of target FT.]{
\begin{tikzpicture}[
    and/.style={and gate US,thick,draw,scale=0.8,fill=black!50,rotate=90,
		anchor=east},
    or/.style={or gate US,thick,draw,scale=0.8,fill=black!20,rotate=90,
		anchor=east},
    be/.style={circle,thick,draw,fill=yellow!90!red,anchor=north,
		minimum width=0.45cm},
    tr/.style={buffer gate US,thick,draw,fill=purple!60,rotate=90,
		anchor=east,minimum width=0.8cm},
    label distance=1mm,
    every label/.style={blue},
    event/.style={rectangle,thick,draw,fill=yellow!20,text width=0.7cm,
		text centered,font=\sffamily,anchor=north},
    edge from parent/.style={very thick,draw=black!70},
    edge from parent path={(\tikzparentnode.south) -- ++(0,-0.4cm)
			-| (\tikzchildnode.north)},
    level 1/.style={sibling distance=3cm,level distance=0.56cm,
			growth parent anchor=south,nodes=event},
    level 2/.style={sibling distance=2cm},
    level 3/.style={sibling distance=1cm},
    level 4/.style={sibling distance=1cm},
    level 5/.style={sibling distance=1cm}
    ]
    \node (g1) [event] {\scriptsize $T$}
    child{
        child{ node(b1) {\scriptsize $B_1$} }
        child{ node(b2) {\scriptsize $B_2$} }
          child{ node(g2) {\scriptsize $G_1$}
            child{ 
              child{ node(b3) {\scriptsize $B_3$} }
              child{ node(b4) {\scriptsize $B_4$} }
            }
            }
            child{ node(g4) {\scriptsize $G_2$}
            child{ 
               child{ node(b5) {\scriptsize $B_5$} }
          	   child{ node(b6) {\scriptsize $B_6$} }
            }
            }
    }
        ;
   \node [or]	at (g1.south)	[]	{};
   \node [and]	at (g2.south)	[]	{};
   \node [and]	at (g4.south)	[]	{};
   \node [be]	at (b1.south)	[]	{};
   \node [be]	at (b2.south)	[]	{};
   \node [be]	at (b3.south)	[]	{};
   \node [be]	at (b4.south)	[]	{};
   \node [be]	at (b5.south)	[]	{};
   \node [be]	at (b6.south)	[]	{};
\end{tikzpicture}
}

\subfloat[Learnt FT.]{
\begin{tikzpicture}[
    and/.style={and gate US,thick,draw,scale=0.8,fill=black!50,rotate=90,
		anchor=east},
    or/.style={or gate US,thick,draw,scale=0.8,fill=black!20,rotate=90,
		anchor=east},
    be/.style={circle,thick,draw,fill=yellow!90!red,anchor=north,
		minimum width=0.45cm},
    tr/.style={buffer gate US,thick,draw,fill=purple!60,rotate=90,
		anchor=east,minimum width=0.8cm},
    label distance=1mm,
    every label/.style={blue},
    event/.style={rectangle,thick,draw,fill=yellow!20,text width=0.7cm,
		text centered,font=\sffamily,anchor=north},
    edge from parent/.style={very thick,draw=black!70},
    edge from parent path={(\tikzparentnode.south) -- ++(0,-0.4cm)
			-| (\tikzchildnode.north)},
    level 1/.style={sibling distance=3cm,level distance=0.56cm,
			growth parent anchor=south,nodes=event},
    level 2/.style={sibling distance=3cm},
    level 3/.style={sibling distance=1cm},
    level 4/.style={sibling distance=1cm},
    level 5/.style={sibling distance=1cm}
    ]
    \node (g1) [event] {\scriptsize $T$}
    child{
        child{ node(g3) {\scriptsize $G_1$}
        child{ 
       		child{ node(b1) {\scriptsize $B_1$} }
         	child{ node(b2) {\scriptsize $B_2$} }
            child{ node(g2) {\scriptsize $G_3$}
            child{ 
              child{ node(b3) {\scriptsize $B_3$} }
              child{ node(b4) {\scriptsize $B_4$} }
            }
            }
        }
        }
     	child{ node(g4) {\scriptsize $G_2$}
            child{ 
               child{ node(b5) {\scriptsize $B_5$} }
          	   child{ node(b6) {\scriptsize $B_6$} }
            }
            }   
    }
        ;
   \node [or]	at (g1.south)	[]	{};
   \node [and]	at (g2.south)	[]	{};
   \node [and]	at (g4.south)	[]	{};
   \node [or]	at (g3.south)	[]	{};
   \node [be]	at (b1.south)	[]	{};
   \node [be]	at (b2.south)	[]	{};
   \node [be]	at (b3.south)	[]	{};
   \node [be]	at (b4.south)	[]	{};
   \node [be]	at (b5.south)	[]	{};
   \node [be]	at (b6.south)	[]	{};
\end{tikzpicture}
}
\subfloat[DNF of learnt FT.]{
\begin{tikzpicture}[
    and/.style={and gate US,thick,draw,scale=0.8,fill=black!50,rotate=90,
		anchor=east},
    or/.style={or gate US,thick,draw,scale=0.8,fill=black!20,rotate=90,
		anchor=east},
    be/.style={circle,thick,draw,fill=yellow!90!red,anchor=north,
		minimum width=0.45cm},
    tr/.style={buffer gate US,thick,draw,fill=purple!60,rotate=90,
		anchor=east,minimum width=0.8cm},
    label distance=1mm,
    every label/.style={blue},
    event/.style={rectangle,thick,draw,fill=yellow!20,text width=0.7cm,
		text centered,font=\sffamily,anchor=north},
    edge from parent/.style={very thick,draw=black!70},
    edge from parent path={(\tikzparentnode.south) -- ++(0,-0.4cm)
			-| (\tikzchildnode.north)},
    level 1/.style={sibling distance=3cm,level distance=0.56cm,
			growth parent anchor=south,nodes=event},
    level 2/.style={sibling distance=2cm},
    level 3/.style={sibling distance=1cm},
    level 4/.style={sibling distance=1cm},
    level 5/.style={sibling distance=1cm}
    ]
    \node (g1) [event] {\scriptsize $T$}
    child{
        child{ node(b1) {\scriptsize $B_1$} }
        child{ node(b2) {\scriptsize $B_2$} }
          child{ node(g2) {\scriptsize $G_1$}
            child{ 
              child{ node(b3) {\scriptsize $B_3$} }
              child{ node(b4) {\scriptsize $B_4$} }
            }
            }
            child{ node(g4) {\scriptsize $G_2$}
            child{ 
               child{ node(b5) {\scriptsize $B_5$} }
          	   child{ node(b6) {\scriptsize $B_6$} }
            }
            }
    }
        ;
   \node [or]	at (g1.south)	[]	{};
   \node [and]	at (g2.south)	[]	{};
   \node [and]	at (g4.south)	[]	{};
   \node [be]	at (b1.south)	[]	{};
   \node [be]	at (b2.south)	[]	{};
   \node [be]	at (b3.south)	[]	{};
   \node [be]	at (b4.south)	[]	{};
   \node [be]	at (b5.south)	[]	{};
   \node [be]	at (b6.south)	[]	{};
\end{tikzpicture}
}
\centering
\caption{Fault Trees may have different equivalent forms.}
\label{fig:learnt-fts}
\end{figure*}

\end{document}